\documentclass[traditabstract]{aa} 
\usepackage{graphicx}
\usepackage{amsmath}
\usepackage{xcolor}
\usepackage[normalem]{ulem}
\usepackage{txfonts}
\usepackage{natbib}
\usepackage{float}
\bibpunct[, ]{(}{)}{;}{a}{}{,}

\begin{document}

\title{Particle-physics constraints from the globular cluster M5:\\
       Neutrino Dipole Moments}

\author{N.~Viaux\inst{1,2,3},
        M.~Catelan\inst{1,2,3},
        P.~B.~Stetson\inst{4},
        G.~G.~Raffelt\inst{5},
        J.~Redondo\inst{5,6},
        A.~A.~R.~Valcarce\inst{7} \and
        A.~Weiss\inst{8}}

   \institute{Pontificia Universidad Cat{\'o}lica de Chile, Departamento de Astronom{\'\i}a y Astrof{\'\i}sica, Av.\ Vicu{\~n}a Mackenna 4860, 782-0436 Macul, Santiago, Chile \email{nviaux@astro.puc.cl}
	     \and
		     Pontificia Universidad Cat\'olica de Chile, Centro de Astroingenier\'ia, Av. Vicu\~na Mackena 4860, 782-0436 Macul, Santiago, Chile
         \and
             The Milky Way Millennium Nucleus, Av.\ Vicu{\~n}a Mackenna 4860, 782-0436 Macul, Santiago, Chile
         \and
             National Research Council, 5071 West Saanich Road, Victoria, BC V9E 2E7, Canada
         \and
             Max-Planck-Institut f\"ur Physik
  (Werner-Heisenberg-Institut),
  F\"ohringer Ring~6, 80805 M\"unchen, Germany
         \and
Arnold Sommerfeld Center, Ludwig-Maximilians-University, Theresienstr. 37, 80333 M\"unchen, Germany
	\and
             Universidade Federal do Rio Grande do Norte, Depto. de F\'{i}sica, 59072-970 Natal, RN, Brazil
         \and
             Max-Planck-Institut f\"ur Astrophysik, Karl-Schwarzschild-Str. 1,
         85748 Garching, Germany
             }

   \date{Received  nn-nn-nn; accepted nn-nn-nn}

\abstract{Stellar evolution is modified if energy is lost in a
``dark channel'' similar to neutrino emission. Comparing modified
stellar evolution sequences with observations provides some of the
most restrictive limits on axions and other hypothetical low-mass
particles and on non-standard neutrino properties. In particular, a
putative neutrino magnetic dipole moment $\mu_\nu$ enhances the
plasmon decay process, postpones helium ignition in low-mass stars,
and therefore extends the red-giant branch (RGB) in globular
clusters (GCs). The brightness of the tip of the RGB (TRGB) remains
the most sensitive probe for $\mu_\nu$ and we revisit this argument
from a modern perspective. Based on a large set of archival
observations, we provide high-precision photometry for the Galactic
GC M5 (NGC~5904) and carefully determine its TRGB position. On the
theoretical side, we add the extra plasmon decay rate brought about
by $\mu_\nu$ to the Princeton-Goddard-PUC (PGPUC) stellar evolution
code. Different sources of uncertainty are critically examined. The
main source of systematic uncertainty is the bolometric correction
and the main statistical uncertainty derives from the distance
modulus based on main-sequence fitting. (Other measures of distance,
e.g., the brightness of RR Lyrae stars, are influenced by the energy
loss that we wish to constrain.) The statistical uncertainty of the
TRGB position relative to the brightest RGB star is less important
because the RGB is well populated. We infer an absolute $I$-band
brightness of $M_I=-4.17\pm0.13$~mag for the TRGB compared with the
theoretical prediction of $-3.99\pm0.07$~mag, in reasonable
agreement with each other. A significant brightness increase caused
by neutrino dipole moments is constrained such that $\mu_{\nu} <
2.6\times 10^{-12}\mu_{\rm B}$ (68\%~CL), where $\mu_{\rm B} \equiv
e/2m_{e}$ is the Bohr magneton, and $\mu_{\nu} < 4.5 \times
10^{-12}\mu_{\rm B}$ (95\%~CL). In these results, statistical
and systematic errors have been combined in quadrature.}

\keywords{globular clusters: general -- globular clusters:
individual: M5 (NGC~5904) --
                Stars: evolution -- Stars: interiors -- Hertzsprung-Russell and C-M diagrams --
                Neutrinos}

\titlerunning{Particle-physics constraints from the globular cluster M5}
\authorrunning{N.~Viaux et al.}

\maketitle


\section{Introduction}\label{sec:intro}

Stellar evolution theory is one of the best established and most
successful theories in the history of astrophysics. Indeed, the
level of agreement between the predictions of canonical stellar
evolution models and observations of globular clusters (GCs), in
particular their color-magnitude diagrams (CMDs), is remarkable
~\citep[e.g.,][]{V00,M09}. However, this excellent level of
agreement can be spoiled if we complement standard evolution with
effects caused by physics beyond the standard model, such as new
weakly interacting particles. As such, GCs can be considered as one
of the largest available laboratories for particle physics studies
\citep{R96}.

The inclusion of a new energy-loss channel implies that the
effective rate of local energy production of the star, $\epsilon =
\epsilon _{\rm nuc}+\epsilon _{\rm grav}-\epsilon _{\nu}$, will
decrease according to $\epsilon = \epsilon _{\rm nuc}+\epsilon_{\rm
grav}-\epsilon _{\nu}-\epsilon _{x}$, where $\epsilon _{x}$ is the
effective rate of local energy production in the new channel. Due to
the additional cooling, a red giant branch (RGB) star will be more
luminous, and have a more massive degenerate He core, at the onset
of helium ignition, than would be the case otherwise, thus giving
rise to a brighter RGB tip \citep[e.g.,][]{sg78,RW92}. The more
massive He core also implies that the horizontal branch (HB) level
will be more luminous (e.g., \citeauthor{sw76} \citeyear{sw76};
\citeauthor{R90} \citeyear{R90}; \citeauthor{R96} \citeyear{R96};
\citeauthor{Ca96} \citeyear{Ca96}).

Neutrino emission becomes more efficient as stars evolve. On the
main sequence (core H-burning), neutrinos are primarily emitted by
nuclear reactions of the proton-proton chain and CNO cycle. In
advanced evolutionary phases, thermal processes dominate, notably
plasmon decay, Compton processes, pair annihilation and
bremsstrahlung. In low-mass stars, near the tip of the RGB (TRGB),
temperatures and densities around $10^{8}\, {\rm K}$ and $10^{6}
{\rm g \, cm^{-3}}$, respectively, are reached in the degenerate
helium core \citep[e.g.,][]{scea07,mc09}. Under these conditions,
plasmon decay $\gamma _{\ast} \rightarrow \nu \overline{\nu}$ is the
most important neutrino loss mechanism. It will be enhanced if
neutrinos have hypothetical direct electromagnetic interactions
caused by milli-charges or dipole moments \citep{b1963}. Such
enhanced neutrino losses are particularly large when the plasma
frequency is large and thus especially important during the RGB,
asymptotic giant branch (AGB), and white dwarf (WD) phases.

One particularly sensitive observable to constrain enhanced energy
losses is the brightness of the TRGB in GCs (\citeauthor{R90}
\citeyear{R90}, \citeauthor{RW92} \citeyear{RW92}, \citeauthor{Ca96}
\citeyear{Ca96}). Together with other observables such as the HB
brightness in several galactic GCs, it was found that the core mass
at He ignition should not exceed its standard value by about
$0.025\,M_\odot$ or 5\% of its standard value,  which in turn means
that the full plasmon decay rate should not exceed its standard
value \citep{H94} by more than a factor of~3. These results
translate into a limit on the neutrino dipole moment of about
$\mu_{\nu} < 3 \times 10^{-12}\mu_{\rm B}$, where $\mu_{\rm B}
\equiv e/(2m_{e})$ is the Bohr magneton, much more stringent than
corresponding laboratory limits (\citeauthor{RPP2012}
\citeyear{RPP2012}). Similar arguments have been used to constrain
novel energy losses by other novel low-mass particles, including
axions, milli-charged particles, Kaluza-Klein gravitons and others
\citep[for a review see][]{Raffelt1999}.

Most of these tests of novel particle physics have not benefited
from the newly available, exquisite CMDs that have become available
only recently, both based on ground- and space-based observations
\citep[e.g.,][]{bs09,ikea12}. These previous tests were also
performed before the revolution in opacity tables and equations of
state (EOS) of the mid-1990s and therefore relied on older tables
\citep{SW73,Cox70,hub69}. In addition, the analysis of systematic
and statistical errors was not detailed enough to assign a clear
quantitative confidence level to the derived constraints, making it
difficult to compare with laboratory results.

The main goal of our work is accordingly to use state-of-the-art
astronomical observations and stellar evolution codes, in order to
obtain new constraints on novel properties of particle physics, with
a more objective assessment of the confidence level on the derived
constraints. Our main result, a new limit on $\mu_\nu$, will be
similar to previous astrophysical limits if the latter are
interpreted as $1\sigma$ constraints. However, our new result is
based on homogeneous observations of a single GC and based on a
detailed error budget.

To achieve this goal, we first provide in Sect.~\ref{sec:obs} a new
CMD for the GC selected for this study, namely M5 (NGC~5904), based
on archival observations. We then briefly describe the stellar
evolution code used in our theoretical calculations
(Sect.~\ref{sec:pgpuc}). The errors due to observational and
theoretical sources are discussed in Sects.~\ref{sec:obs-unc} and
\ref{sec:teo-unc}. 
In Sect.~\ref{sec:comp} we compare observations with 
theoretical models to set constraints on $\mu_\nu$. Our conclusions are drawn in 
Sect.~\ref{sec:conclusions}. In future papers, we plan to extend
our analysis to other GCs and consider other particle properties,
notably those of the hypothetical axion.

\section{Observational framework}\label{sec:obs}

\subsection{Cluster selection}

In principle, each of the fully resolved and well-populated Milky
Way GCs could be useful for our study. However, to derive the most
meaningful particle-physics constraints, we impose stringent
criteria on the properties of the selected GC, providing us with
clean CMDs and straightforward interpretation of the results. First,
we require that the cluster be sufficiently massive, with an
absolute magnitude satisfying $M_V < -8.0$~mag, so as to ensure a
sufficiently well-populated CMD.\footnote{Unless otherwise noted, GC
parameters are adopted from the Feb.~2010 version of the
\citet{wh96} catalog.} Second, we restrict the amount of foreground
reddening to a maximum of $E(B-V) = 0.1$~mag, which also helps
reduce the possibility of significant differential reddening towards
the cluster. Third, we make sure that the cluster metallicity is
neither too high nor too low, leading to a fairly uniformly
populated HB, thus avoiding GCs with extremely red or blue HB
morphologies. Fourth, candidates must be sufficiently close that
deep, high-quality photometric data are available. Last but not
least, we avoid GCs for which there is strong evidence of multiple
CMD sequences, and in particular, those for which the level of
internal variation in the He abundance has been suggested to be high
\citep{rgea00}. Note that the TRGB is unaffected from the O-Na
anomaly that is present in GCs \citep{salaris06}.

This leaves us with a much reduced set of GCs, with M5 at the top of
our list. This is a well-studied, fairly massive cluster, with $M_V
= -8.81$~mag, a moderate metallicity of ${\rm [Fe/H]} = -1.29$, and
a foreground reddening of only $E(B-V) = 0.03$~mag. The distance
modulus of the cluster is~\citep{L05}
\begin{equation}
\label{distance}
(m-M)_0=14.45 \pm 0.11
\end{equation}
which corresponds to a modest distance of 7.5~kpc from the Sun. The
cluster distance to the Galactic center is about
6.2~kpc.\footnote{M5 actually appears to be an outer-halo GC that
spends much of its time at distances larger than $\sim 50$~kpc, but
which just happens to lie close to its perigalacticon at this point
in time \citep[][and references therein]{kc97}.}

Most importantly for our purposes, the recent, detailed
spectroscopic and photometric study by \citet{gratton12} reveals no
evidence for a significant internal spread in the He abundance,
except perhaps for the component of the cluster that falls on the
extreme blue end of its HB --- which, according to the analysis by
\citet{rgea00}, could be enhanced in He by a maximum of $\Delta Y
\approx 0.03$. This is to be compared with much higher levels of He
enhancement suggested for other GCs, such as NGC~2808 and
$\omega$~Centauri~= NGC~5139, which may reach $\Delta Y \approx
0.14$ \citep[e.g.,][and references therein]{ikea12,amea12}.

Accordingly, we will focus on M5, and in future extend our study to
other GCs in the reduced sample just described.

\subsection{Data acquisition and photometry}\label{sec:phot}

We have carried out crowded-field, point-spread function (PSF)
photometry for M5 using the DAOPHOT II/ALLFRAME suite of programs
\citep{pbs87,pbs94}. For this purpose we have amassed an extensive
database comprised of data sets obtained from many different
sources, including public archives. This effort represents a
continuation of previous similar work that was carried out for other
GCs \citep[see][for a review and references]{pbs09}.

The current corpus of observations for M5 consists of 2840 CCD
images obtained during 40 observing runs on 12 telescopes over a
span of 27 years. Details of the observations are provided in
Table~\ref{obs}. The measurements were contained in 147 separate
data sets, each of which was individually transformed to the
photometric system of \citet{l92} as described elsewhere
\citep[e.g.,][]{st2000,st2005}. In the case of the M5 observations,
114 of the data sets were reduced in photometric (all-sky) mode,
while the remaining 33 were reduced in non-photometric (local) mode.

\begin{table*}
\caption{NGC~5904 Observations} \label{obs}
\centerline{\begin{tabular}{r l l l l c c c c c c}     
\hline\hline
 &   &     &  &  &   &   &  Exposures &   &   &  \\
\# & Run ID  &   UT Dates  & Telescope & Instrument &  U &  B &  V &  R &  I & Multiplex \\
\hline
   1  &  ct84   &   1984 03 09    &  CTIO 4.0m   &  RCA     &     - &  2 &  2 &  - & -&- \\
   2  &  cf84   &   1984 06 24    &  CFHT 3.6m   &  RCA1    &     - &  1 &  1 &  - & -&- \\
   3  &  ct85   &   1985 04 18    &  CTIO 4.0m   &  RCA     &    -  & 1  & 1  & -  & -&- \\
   4  &  kp36   &   1985 06 13-15 &  KPNO 0.9m   &  RCA     &     - &  3 &  6 &  - & -&- \\
   6  &  jvw    &   1986 03 28-30 &  INT 2.5m    &  RCA     &     3 &  4 &  7 &  4 & 4&- \\
   7  &  f1     &   1987 03 10    &  CTIO 4.0m   &  RCA     &     4 &  5 &  7 &  - & 8&- \\
   8  &  ct87   &   1987 06 27-29 &  CTIO 4.0m   &  RCA     &     - &  2 &  2 &  - & -&- \\
   9  &  ct88   &   1988 06 13-16 &  CTIO 4.0m   &  RCA5    &     - &  7 &  7 &  - & -&- \\
  10  &  bol    &   1988 06 15-16 &  CTIO 0.9m   &  RCA4    &     - & 18 & 40 &  - & 37&-\\
  11  &  bol89  &   1989 08 29-30 &  CTIO 4.0m   &  TI2     &     - &  3 &  3 &  - & - &-\\
  12  &  pac    &   1990 06 26    &  TNG 3.6m    &  GEC6    &     6 & 10 &  - &  - & - &-\\
  13  &  cf91   &   1991 07 05-07 & CFHT 3.6m    & Lick2    &    -  & 2  & 2  & -  & - &-\\
  14  &  saic   &   1992 05 25    &  CFHT 3.6m   &  HRCam/saic1 &  - & - &  8 & 13 & - &-\\
  15  &  cf92   &   1992 06 09    &  CFHT 3.6m   &  RCA4    &     -  & 3 &  - &  - & - &-\\
  16  &  emmi4  &   1992 07 05    &  ESO NTT 3.6m&  EMMI    &     -  & - &  6 &  - & 6 &-\\
  17  &  bol93  &   1993 06 17-20 &  CTIO 0.9m   &  Tek1K-1 &     -  & 8 & 11 &  - & 17&-\\
  18  &  bolte  &   1994 04 14-16 &  KPNO 2.1m   &  t1ka    &     -  & 7 & 26 &  - & 20&-\\
  19  &  pwm    &   1994 04 21-27 &  JKT 1.0m    &  EEV7    &     -  & 11&  62&  - & 65&-\\
  20  &  siv    &   1994 05 01    &  INT 2.5m    &  EEV5    &     -  & - & 21 &  - & 21&-\\
  21  &  cf94   &   1994 06 05-06 &  CFHT 3.6m   &  Loral3  &     -  & 19& -  &  - & 25&-\\
  22  &  fgj96  &   1996 06 03    & NOT 2.6m     & CCD      &    -   & 1 &  1 &  - & 1 &-\\
  23  &  bond9  &   1997 05 08    &  KPNO 0.9m   &  t2ka    &     3  & 3 &  3 &  - & 3 &-\\
  24  &  arg    &   1997 06 01    &  JKT 1.0m    &  TEK2    &     -  & - &  3 &  - & 3 &-\\
  25  &  bond11 &   1998 03 22    &  KPNO 0.9m   &  t2ka    &     4  & 4 &  4 &  - & 4 &-\\
  26  &  bond6  &   1998 04 18    &  CTIO 0.9m   &  Tek2K\_3 &     1  & 1 &  1 &  - & 1 &-\\
  27  &  dmd    &   1998 06 24    &  JKT 1.0m    &  TEK4    &     -  & - &  2 &  - & 2 &-\\
  28  &  jkt9905&   1999 05 13    &  JKT 1.0m    &  TEK4    &     -  & 2 &  2 &  2 & - &-\\
  29  &  wfi12  &   1999 07 06-08 &  MPI/ESO 2.2m&  WFI     &     -  & - & 35 &  - & 40&8\\
  30  &  wfi4   &   2000 04 21-23 &  MPI/ESO 2.2m&  WFI     &     7  & - & 12 & 10 & 10&8\\
  31  &  benetti&   2000 04 28    & TNG 3.6m     & OIG      &    -   & 2 &  2 &  - & 4 &2\\
  32  &  wfi10  &   2000 07 08-12 &  MPI/ESO 2.2m&  WFI     &     -  & 2 &  3 &  - & - &8\\
  33  &  wfia   &   2000 04 22-26 &  MPI/ESO 2.2m&  WFI     &    16  & 19&  20& 26 & 25&8\\
  34  &  wfib   &   2000 07 11-12 &  MPI/ESO 2.2m&  WFI     &    16  & 19&  20& 26 & 25&8\\
  35  &  not017 &   2001 07 09-16 &  NOT 2.6m    &  CCD7    &     -  & 9 & 11 &  - & 10&-\\
  37  &  hannah &   2002 03 29    &  JKT 1.0m    &  SIT2    &     -  & 5 &  5 &  5 & - &-\\
  39  &  soar10 &   2010 05 12    &  SOAR 4.1m   &  SOI     &     5  & 6 &  6 &  - & 10&2\\
  40  &  int11  &   2011 08 21    &  INT 2.5m    &  WFC     v     1  & 1 &  1 &  - & 1 &4\\
\hline
\end{tabular}}

\vspace{0.5cm}
Notes:
 1.~Observers J.~Hesser, R.~McClure.
 2.~Observers J.~Hesser, R.~McClure.
 3.~Observers J.~Hesser, R.~McClure.
 4.~Observer P.~Stetson.
 5.~Observers P.~Stetson, W.~Harris.
 6.~Observer ``JVW''.
 7.~Observer N.~Suntzeff.
 8.~Observer P.~Stetson.
 9.~Observer P.~Stetson.
10.~Observer M.~Bolte. 11.~Observer M.~Bolte. 12.~Observer ``PAC''.
13.~Observer P.~Stetson. 14.~Observers M.~Bolte, J.~Hesser.
15.~Observer P.~Stetson. 16.~Program identification unknown,
observer unknown. 17.~Observer M.~Bolte. 18.~Observer M.~Bolte.
19.~Observer ``PWM''. 20.~Observer ``SIV''. 21.~Observers
P.~Stetson, D.~VandenBerg. 22.~Observer F.~Grundahl. 23.~Observer
H.~Bond. 24.~Observer A.~Rosenberg. 25.~Observer H.~Bond.
26.~Observer H.~Bond. 27.~Observer ``DMD''. 28.~Observer unknown.
29.~Program identification unknown, observer unknown. 30.~Program
identification unknown, observer Momany. 31.~Observer Benetti.
32.~Program identification 065.L-0463, observer Ferraro. 33.~Program
identification 065.O-0530, observer Momany. 34.~Program
identification unknown, observer Altavilla. 35.~Observer H.~Bruntt.
36.~Program identification 68.D-0265(A), observer unknown.
37.~Observer J.~Mendez. 38.~Program identification 71.D-0220(A),
observer unknown. 39.~Proposal ID 2010A-0160, observers Kuehn,
Smith, Catelan. 40.~Proposal C98, observer Milone.
\end{table*}

It should be noted that detectors of different projected areas were
directed at various pointings in the cluster. Furthermore, each of
the Optical Imager Galileo on Telescopio Nazionale Galileo and the
SOAR Optical Imager on the Southern Astrophysical Research telescope
consists of a pair of non-overlapping CCDs; the Wide Field Camera on
the Isaac Newton Telescope contains four detectors; and the Wide
Field Imager on the Max Planck Gesellschaft/European Southern
Observatory telescope contains eight (``Multiplex'' column in
Table~\ref{obs}). As a result, no individual star appeared in all
the images. In fact, the greatest number of independent observations
for any star was 34 in the $U$ band, 77 in $B$, 134 in $V$, 35 in
$R$, and 128 in $I$.

\begin{figure}
\centering
\includegraphics[width=0.9\columnwidth]{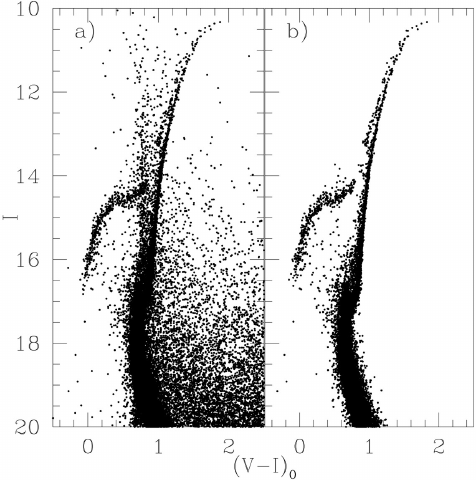}
\caption{Color-magnitude diagram of M5. {\em Left}: Original.
{\em Right}: After field star decontamination.}
\label{cmdM5}
\end{figure}

Figure~\ref{cmdM5}a shows the derived CMD which is clearly
contaminated with field stars. We remove the field component with a
statistical decontamination, using the method described in
\citet{G03}. The cleaned CMD is shown in Fig.~\ref{cmdM5}b.

Details and uncertainties regarding the calibration and the
possibility of saturation close to the TRGB will be discussed
further in Sect.~\ref{sec:obs-unc}.
\newpage

\subsection{Finding the TRGB}\label{sec:trgb}
\subsubsection{RGB and AGB stars in M5}

One of our main points of interest is the TRGB. For Milky Way GCs it
is not trivial to identify this point in the CMD because the
evolution close to the TRGB is fast, and accordingly this region of
the CMD is thinly populated. The brightest RGB star, apart from
small photometric errors, provides a lower limit to the brightness
of the TRGB. Therefore, one task is to identify the brightest RGB
star in M5. Moreover, we need to estimate the likely brightness
difference $\Delta_{\rm tip}= I_1-I_{\rm TRGB}$ between the TRGB
($I_{\rm TRGB}$) and the brightest star ($I_1$) in the I-band, or
rather, the probability distribution for $\Delta_{\rm tip}$ in view
of the overall RGB population.

\begin{figure}[h]
  \centering
  \includegraphics[width=0.8\columnwidth]{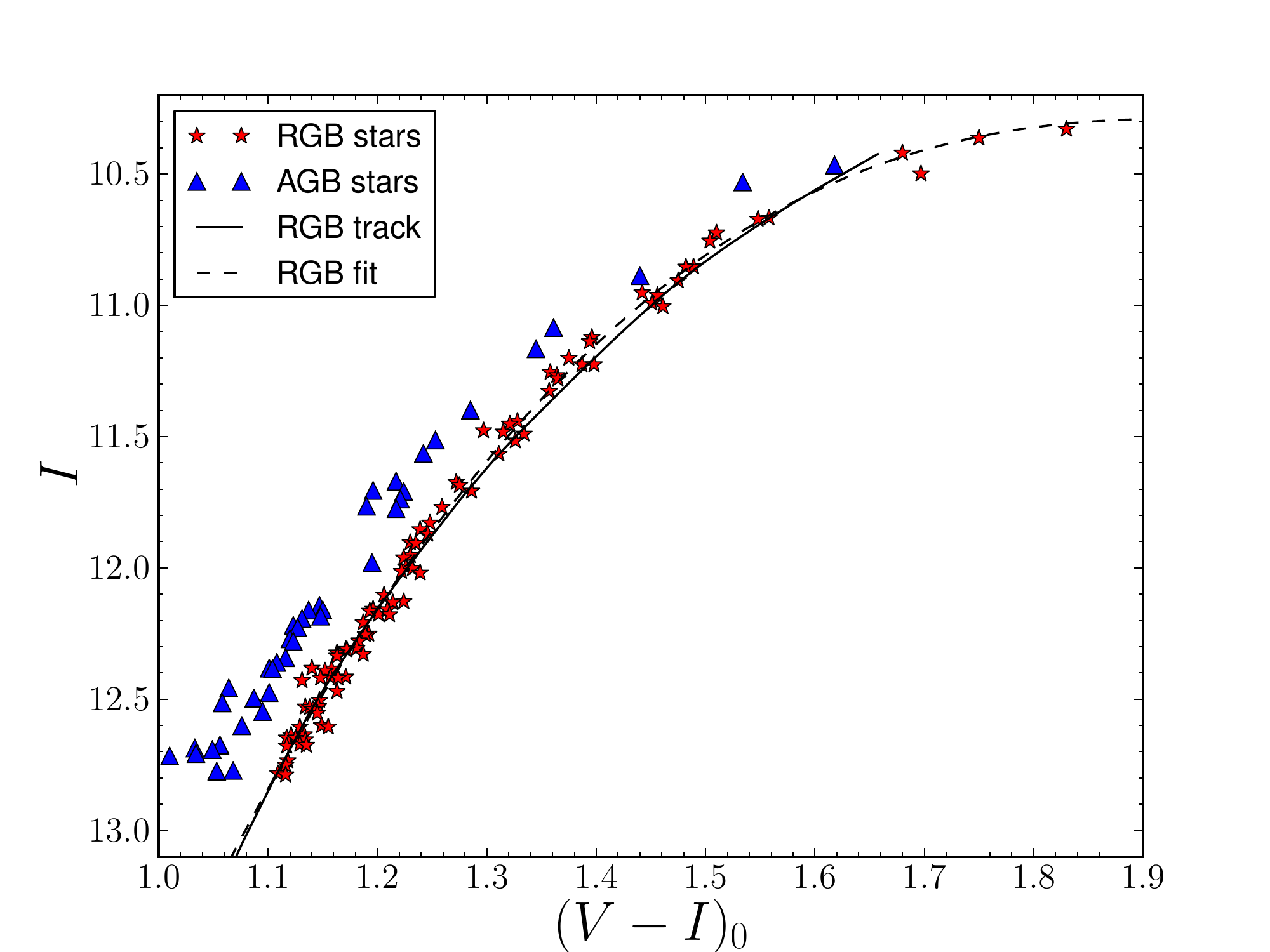}
  \caption{Upper part of the CMD for M5.
  Nominal identification of RGB and AGB stars as explained in the text. Solid line: RGB track with initial parameters described in \S 3.2. Dashed line: Nominal
  RGB according to the fit function of Eq.~(\ref{eq:RGBfit}).
  \label{fig:cmdRGB}}
\end{figure}

In order to evaluate this statistical distribution we first show, in 
Fig.~\ref{fig:cmdRGB}, the upper CMD for M5. The RGB and AGB are well separated and
we assign nominal membership to one of these branches based on the
distance of a given star from the RGB. As a first possibility to
describe the locus of the RGB we have tried an isochrone which here
is identical to the evolutionary track of a star with our benchmark
parameters (see the solid line in Fig.~\ref{fig:cmdRGB}. However, it does not
provide a good fit near the TRGB. We prefer to use a simple
empirical fit function of the form
\begin{equation}\label{eq:RGBfit}
I=I_0+3.83\,[1.95 - (V-I)]^{2.5}
\end{equation}
shown as a solid line. We use $I_0=I_1-0.04~{\rm mag}=10.289$~mag
where $I_1=10.329$~mag refers to the brightest star

The crest value $I_0$  of our fit function is 0.04~mag brighter than
the brightest star and a reasonable first estimate for $I_{\rm
TRGB}$. Therefore, we think of $I_0$ as our provisional TRGB and
study the distribution of stars relative to this reference point. We
cut the CMD at 2.5~mag below $I_0$, safely away from the HB and the
RGB bump, yet including as many as 136 stars. Had we used the
brightest star as a reference point, a 2.5~mag interval would have
reached 0.04~mag dimmer, including 5 more stars. In other words, our
somewhat arbitrary choice of reference point has only a minor impact
on the overall statistics.

\begin{figure}[t]
  \centering
  \includegraphics[width=0.7\columnwidth]{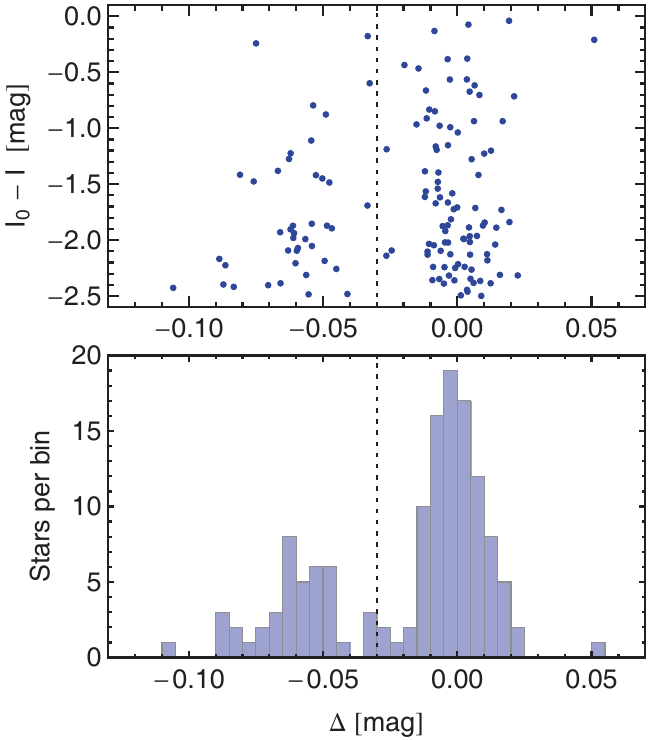}
  \caption{Distance $\Delta$ from our RGB reference line
  of Eq.~(\ref{eq:RGBfit}) and distribution of stars
  as a function of $\Delta$.
  \label{fig:Delta}}
\end{figure}

We next determine the distance $\Delta$ of every star in our
remaining ensemble relative to our reference line. In
Fig.~\ref{fig:Delta} we show the star distribution as a function of
$\Delta$. Note that the horizontal scale in Fig.~\ref{fig:Delta} (in
magnitudes) is strongly expanded relative to the vertical scale,
i.e., the RGB is almost vertical. In this sense, $\Delta$ is
essentially the color difference to the reference line, except near
the TRGB, where this line becomes almost horizontal.

Our reference line tracks the RGB very well indeed and the AGB and
RGB are well separated. Stars with $\Delta<-0.04$ are almost
certainly on the AGB whereas those with $-0.02<\Delta$ are almost
certainly on the RGB. In addition, there are six stars almost lined
up at $\Delta=-0.03$ (dashed line) where the association with one of
the branches is not obvious. We use the dashed line as a formal
separation---in this way these six stars are equally divided. With
this choice, we find $N_{\rm RGB}=95$ and $N_{\rm AGB}=41$.

\begin{figure}[t]
  \centering
  \includegraphics[width=0.8\columnwidth]{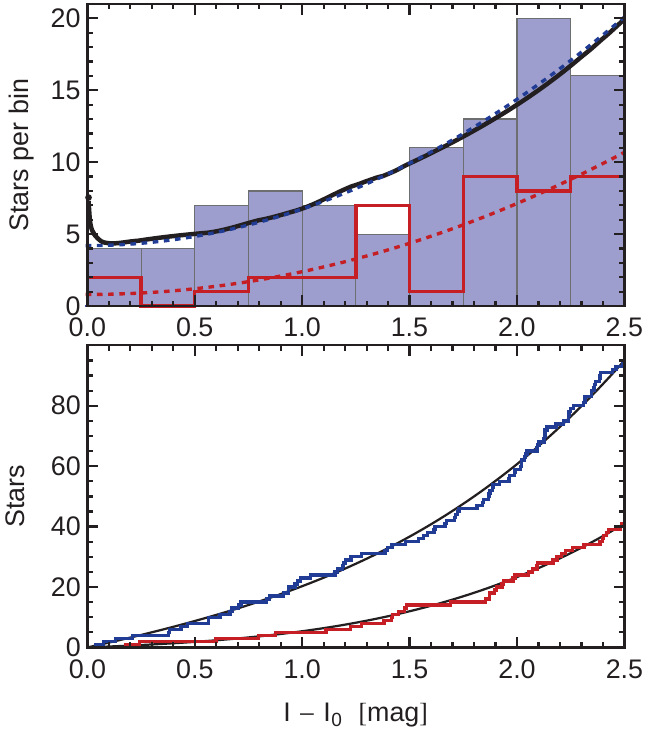}
  \caption{{\it Top:} Luminosity functions of RGB (filled histogram) and AGB.
  Thick black line: Expectation from evolutionary speed. Dashed lines: Fit to data.
  {\it Bottom:} Cumulative luminosity function for RGB stars (in blue) and AGB stars (in red) and fit functions for each case.
  \label{fig:distribution}}
\end{figure}

Next we show the $I$-brightness distribution of our RGB and AGB
stars in Fig.~\ref{fig:distribution}. From an evolutionary RGB track
(described in Sect. \ref{sec:pgpuc1}) we can predict this
distribution from the evolutionary speed $(dI/dt)^{-1}$ shown as a
black thick solid line in the upper panel. The evolution accelerates
as the star ascends the RGB, corresponding to fewer bright stars.
Just before reaching the TRGB, the brightness increase slows down,
reaches a peak, and then quickly drops. Therefore, $(dI/dt)^{-1}$
has a cusp at the TRGB.

In our brightness range, and ignoring this cusp, the theoretical
distribution is rather well represented by a function of the form
\begin{equation}\label{eq:distributionfit}
\frac{dN_{\rm RGB}}{dI}=N_{\rm RGB}\,\left[a + b\,(I-I_0)^2\right]\,,
\end{equation}
shown as a dashed line. The corresponding cumulative distribution
looks even more similar to the one from the numerical evolution
speed. The requirement that the distribution is normalized to the
total number of stars in our 2.5~mag brightness interval relates the
coefficients as $b=12\,(2-5 a)/125$ and numerically we find
$a=0.177$ and $b=0.107$.

For the AGB, we do not have a good theoretical prediction because
the evolution is strongly affected by mass loss. A function of the
form of Eq.~(\ref{eq:distributionfit}) with $a=0.0787$ and $b=0.154$
is a reasonable fit (Fig.~\ref{fig:distribution}). Our overall
number of AGB stars is roughly half of the RGB, but the AGB stars
die out more quickly toward the TRGB. In the 1~mag brightest
interval we have 5 AGB stars (including two that might be
questionable) and 23 RGB stars, i.e., the random chance that a star
in the top brightness range is an AGB star is roughly 18\%. This
approximate fraction also derives from our fit parameters $a$ and
the total number of stars as $a_{\rm AGB}N_{\rm AGB}/(a_{\rm AGB}
N_{\rm AGB}+a_{\rm RGB} N_{\rm RGB}) =16\%$.

\subsubsection{TRGB relative to brightest star}

In order to derive a probability distribution for $\Delta_{\rm
TRGB}= I_1-I_{\rm tip}$ we have used the results of the previous
section and have performed Monte Carlo simulations of the star
distribution in the $I$-band. We use 95 RGB stars in our 2.5~mag
fiducial interval, following on average the theoretical
distribution, i.e., the thick black line in
Fig.~\ref{fig:distribution}. The observed distribution in M5 agrees
very well with this expectation, i.e., it looks like a typical
random realization of the expected distribution.

\begin{figure}
\centering
\includegraphics[width=0.8\columnwidth]{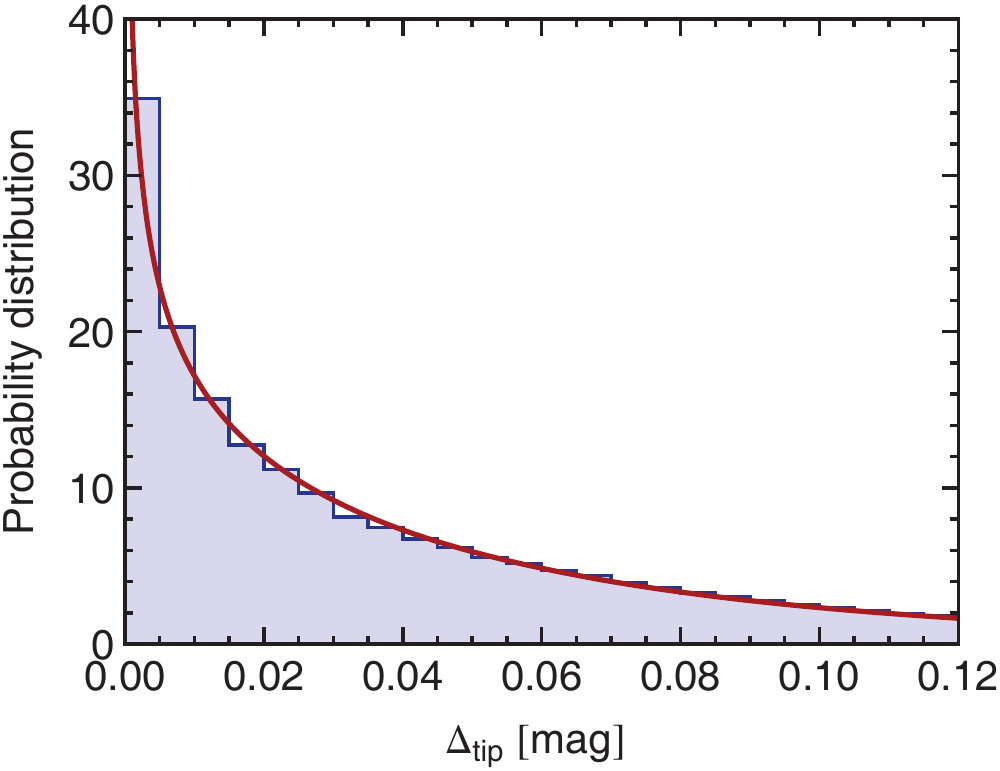}
\caption{Probability distribution for $\Delta_{\rm tip}=I_1-I_{\rm TRGB}$
between the TRGB and the brightest star. The histogram is from
a Monte Carlo simulation, the red line is the analytic
representation of Eq.~(\ref{eq:probfit}).\label{fig:probfit}}
\end{figure}

In Fig.~\ref{fig:probfit} we show the $\Delta_{\rm tip}$
distribution from a large number of Monte Carlo realizations. It is
well approximated by an exponential plus a cusp, reflecting the cusp
of $(dI/dt)^{-1}$ near the TRGB. A good analytic representation,
normalized to unity, is
\begin{equation}\label{eq:probfit}
p(\Delta_{\rm tip})=\frac{e^{-\Delta_{\rm tip}/\lambda}}{\lambda}
\,\frac{1+a\,\Delta_{\rm tip}^{-1/3}}{1+a\,\lambda^{-1/3}\,\Gamma_{2/3}}\,,
\end{equation}
where $\Gamma_{2/3}\approx1.3541$ is the Gamma function at argument
$2/3$, $\lambda=0.068~{\rm mag}$ represents the exponential decline,
and $a=2.30\, {\rm mag}^{1/3}$ the cusp contribution.

Additional neutrino cooling caused by $\mu_\nu$ shifts the TRGB to
brighter values and in this way changes the expected brightness
distribution relative to the TRGB. Repeating our exercise with
evolutionary tracks for the relevant range of $\mu_\nu$ values, we
find only small modifications of the implied parameters $\lambda$
and $a$. Likewise, we have repeated this exercise for other BC
prescriptions and also find only a small influence.

We conclude that in M5, almost independently of $\mu_\nu$ and the BC
used (see Sect. ~\ref{sec:teo-unc}), the TRGB is within 0.05~mag
(68\%~CL), and within 0.16~mag (95\%~CL) of the brightest star.
Moreover, the average value and rms variance are
\begin{equation}\label{eq:statisticalshift}
\langle\Delta_{\rm tip}\rangle=0.048~{\rm mag}
\quad\hbox{and}\quad
\sigma_{\Delta_{\rm tip}}=0.058~{\rm mag}\,,
\end{equation}
respectively. Note that the distribution is very asymmetric and
$\sigma_{\rm tip}$ is the formal rms variance, not a Gaussian error.

\subsubsection{Identifying the brightest RGB star in M5}

What remains is to identify the brightest RGB star in M5. Our
selection based on color suggests that the brightest stars are on
the RGB, although our empirical track was somewhat constructed in
this way. Based on distance from the isochrone (solid line in 
Fig.~\ref{fig:cmdRGB}), however, would put even more of the brightest stars on the
RGB. Still, this assignment may not be entirely trustworthy. Of
course, the paucity of AGB stars alone tells us that it is extremely
unlikely that the brightest stars are all on the AGB. Another
possibility for discrimination is based on chemical abundance
variation. In Table~\ref{table:0} we list the identification of the three
brightest stars by different authors.

\begin{table*}[t]
\caption{Evolutionary state for the brightest stars in M5.}
\label{table:0}
\centering
\begin{tabular}{lllll}
\hline\hline
Brightest star   & \hspace{-0.2cm}1st        & 2nd         & 3rd         & Method
for AGB/RGB separation   \\
I-magnitude      & \hspace{-0.2cm}10.329     & 10.363      & 10.420      &     \\
RA (J2000)       & \hspace{-0.2cm}15 18 36.03& 15 18 34.18 & 15 18 34.18 &     \\
Dec (J2000)      & \hspace{-0.2cm}2 6 38.1   & 2 6 25.9    & 2 6 41.7    &     \\
\multicolumn{5}{l}{Evolutionary state:}\\
--Our selection  &\hspace{-0.2cm}RGB&RGB&RGB & Color\\
--\citet{sand04} & \hspace{-0.2cm}RGB& AGB & RGB & Color\\

--\citet{ivans01}& \hspace{-0.2cm}RGB & RGB&RGB & Chemical composition   \\
\hline
\end{tabular}
\end{table*}

All evidence points to the brightest star in M5 being on the RGB. If
this were not the case after all, we should use the second or third
brightest star. The TRGB would then be found dimmer by 0.034 or
0.091 mag, allowing for less novel particle emission. Therefore,
assuming the brightest star to be on the RGB is conservative for the
purpose of deriving particle-physics limits.

\citet{MF95} tested four possible observational problems (photometric errors, crowding, contamination by field stars, and population size) that can affect the empirical determination of the TRGB in globular clusters and nearby galaxies. Specifically, using bins of 0.1 mag, there is claimed that, in order to have a precision of 0.1 mag, are required a signal-to-noise ratio greater than 5, a crowding with 25\% in the first three magnitudes, no more than 20 field stars per bin, and a minimum of 100 stars in the brightest bin. Even though these are good tests,  observations and decontamination techniques have had huge improvements during the last two decades that CMD of GCs show narrow stellar populations \citep{sand04}. In particular, using present-day observations is possible to define the minimum luminosity of the TRGB with high precision if the identification of the brightest RGB stars is done thoroughly, as is the goal of \S 2.3.2.

\section{Theoretical framework}
\label{sec:pgpuc}

\subsection{Neutrino emission rate}

Throughout this paper we use the Princeton-Goddard-PUC
\citep[PGPUC;][]{pgpuc} code in our evolutionary calculations.
Our benchmark tracks do not implement mass loss on the RGB.
The dominant neutrino emission process on the RGB evolution is
plasmon decay for which PGPUC uses the analytic approximation
formulas of \citet{H94}.

To incorporate $\mu_\nu$ effects, we follow the procedure prescribed
by \citeauthor{RW92} (\citeyear{RW92}), i.e., we scale the standard
plasmon emission rate by the prescription given in their Eqs.~(9)
and (10). The discussion of standard and nonstandard plasma emission
rates of \citet{H94} reveals that the error introduced by this
simple prescription can be around 10\% for the relevant conditions
and is therefore not crucial for deriving bounds on $\mu_\nu$.

\subsection{Modified stellar models}
\label{sec:pgpuc1}
To show the effects of different $\mu_\nu$ values, we compute
evolutionary tracks with an initial mass of $M=0.82\, M_{\odot}$,
$Y=0.245$, $Z=0.00136$, and $[\alpha/{\rm Fe}]=+0.30$ for the range
$\mu_{\nu}=(\hbox{0--9}) \times 10^{-12} \mu_{\rm B}$. The initial
parameters ($M$, $Y$, $Z$, and $[\alpha/{\rm Fe}]$) were chosen
according to the following criteria:
\begin{itemize}
\item For the metallicity we adopt ${\rm [Fe/H]}=-1.33 \pm
    0.02$~dex, from \citet[][see also \citealt{kmcw10}]{Carr09}.
    With an abundance of the $\alpha$-capture elements given by
    $[\alpha/{\rm Fe}]=+0.30$ \citep[][]{dlea11}, this gives a
    global metallicity value of $Z=0.00136$ for M5.\footnote{For
    instance, according to \citet{salaris93}, the global
    metallicity can be computed as ${\rm [M/H]} = {\rm
    [Fe/H]}+\log(0.638\,f_{\alpha} + 0.362)$, where $f_{\alpha}
    = 10^{[\alpha/{\rm Fe}]}$. In our case, we use the
    metallicity calculator that is available in the PGPUC Online
    webpage, {\tt
    http://www2.astro.puc.cl/pgpuc/FeHcalculator.php}.}
\smallskip
\item The He abundance was selected assuming that M5 is
    sufficiently metal-poor that no significant He enhancement
    took place before the formation of its stars. Thus a He
    abundance close to the primordial nucleosynthesis value
    \citep[e.g.,][]{yiea07,gs07,gs12} is applicable, as indeed
    favored by detailed CMD studies of Galactic GCs
    \citep[e.g.,][]{msea04}.
\smallskip
\item The mass of the star at the TRGB ($M=0.82 \, M_{\odot}$)
    corresponds to what is expected for the adopted chemical
    composition and an age similar to the age of the Universe,
    currently estimated at 13.8~Gyr \citep{ekea11}.
\end{itemize}

\begin{figure}[t]
   \centering
   \includegraphics[width=0.8\columnwidth]{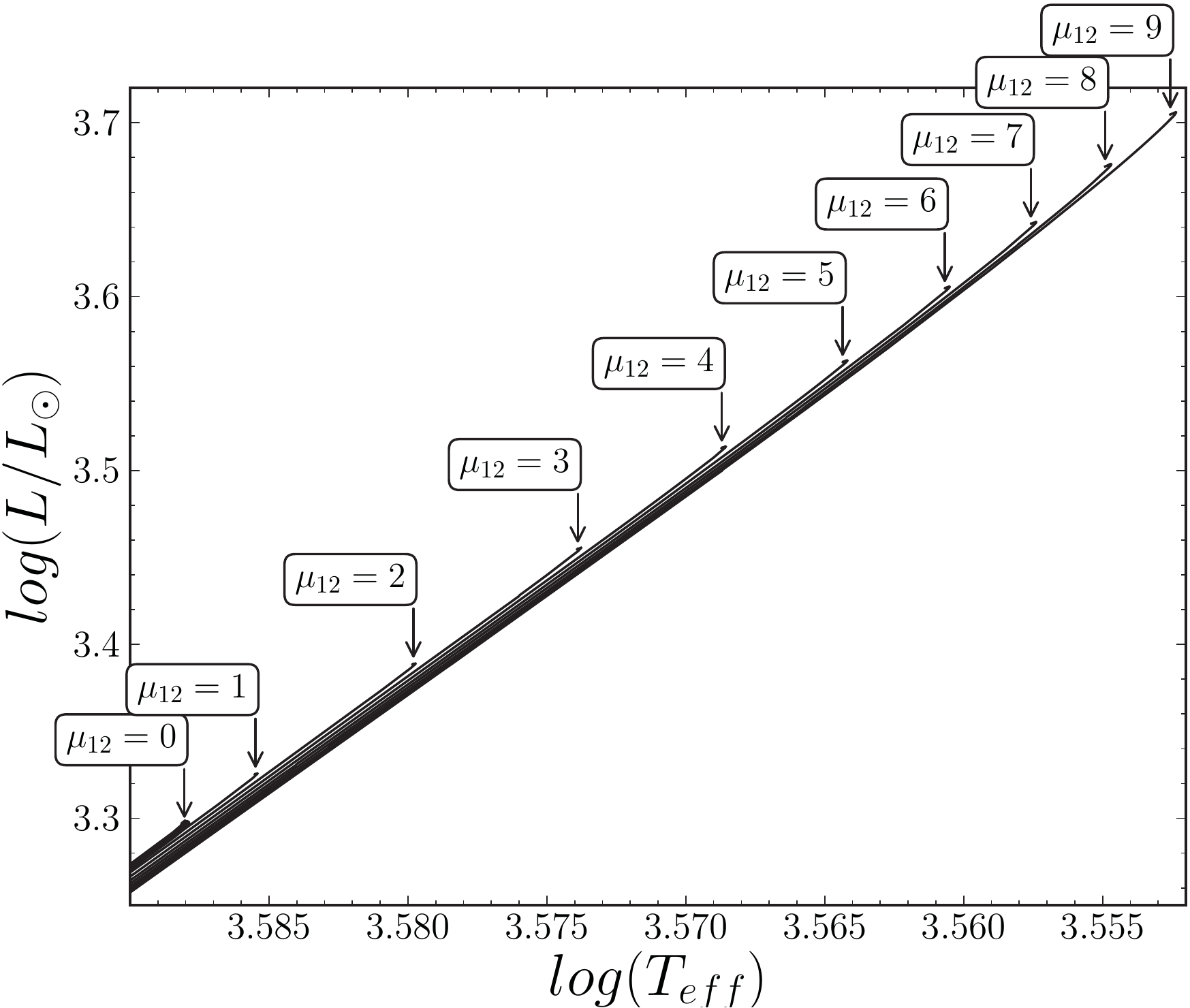}
   \caption{PGPUC evolutionary tracks around the TRGB for the indicated values
   of $\mu_{\nu}$, where
   $\mu_{12}=\mu {_\nu}/10^{-12}\mu_{\rm B}$.}
   \label{hr}
\end{figure}

\begin{figure}[t]
   \centering
   \includegraphics[width=0.8\columnwidth]{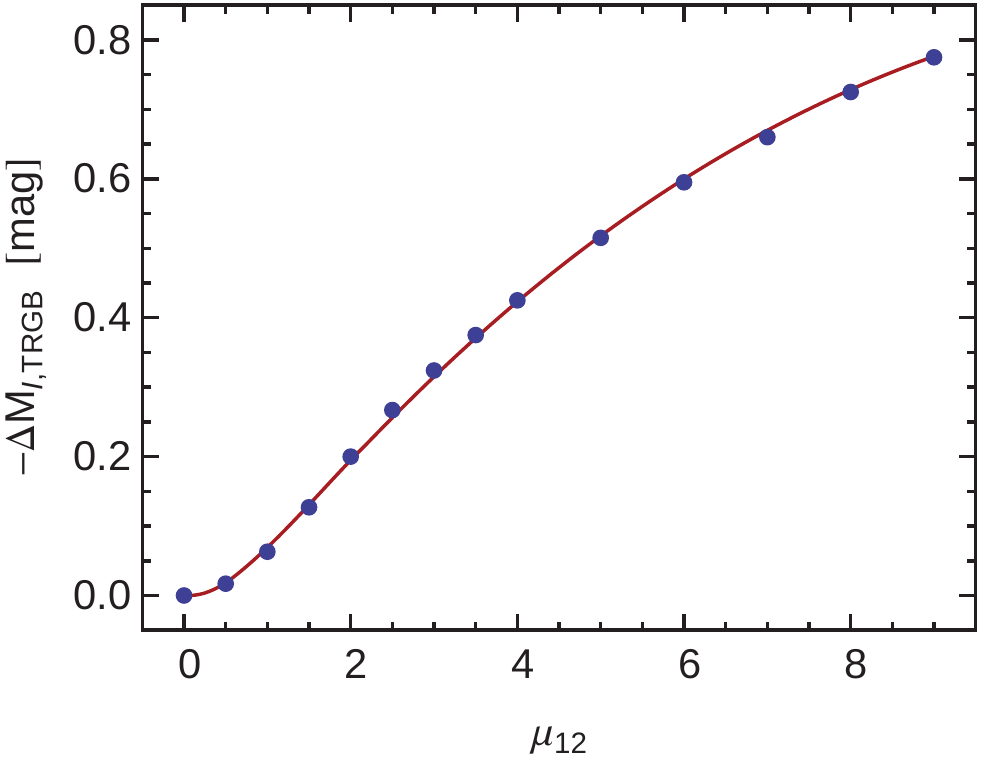}
   \caption{$I$-band brightness increase of the TRGB corresponding to the
   modified tracks shown in Fig.~\ref{hr}. We also show the analytic approximation
   formula of Eq.~(\ref{eq:DeltaMI}).}
   \label{fig:DeltaMI}
\end{figure}

Figure~\ref{hr} shows our evolutionary tracks close to the TRGB,
where the $\mu_\nu$ effects become important. 
For our later comparison with observational data, we transform the 
relative luminosity $L/L_\odot$ into $I$-band absolute brightness 
using the bolometric corrections of \citet{W11}. 
The results, presented in Fig.~\ref{fig:DeltaMI}, can be expressed 
in terms of a simple analytic fit formula, 
\begin{equation}\label{eq:DeltaMI}
M^0_{I,{\rm TRGB}}=-4.03-0.23\,\left(\sqrt{\mu_{12}^2+0.64}-0.80-0.18\,\mu_{12}^{1.5}\right)\,,
\end{equation}
where $\mu_{12}=\mu {_\nu}/10^{-12}\mu_{\rm B}$.

As explained in the earlier literature (\citeauthor{R90}
\citeyear{R90}, \citeauthor{RW92} \citeyear{RW92}, \citeauthor{Ca96}
\citeyear{Ca96}), the dominant effect of increased neutrino losses
on the upper RGB is to allow the degenerate He core to grow to a
somewhat larger mass. In this sense the effects of novel energy
losses can be expressed in terms of the core mass at He ignition
which, in turn, determines the brightness at the TRGB. We here
prefer to discuss our results directly in terms of $\mu_\nu$, but
for comparison we provide the core-mass increase in terms of an
analytic fit formula
\begin{equation}\label{eq:coremass}
\Delta M_{\rm core}=0.023\,M_\odot\,
\left[\sqrt{\mu_{12}^2+1.05^2}-1.05-0.16\,\mu_{12}^{1.5}\right]\,.
\end{equation}
This result is very similar to Eq.~(11) of \citeauthor{RW92}
(\citeyear{RW92}) in spite of the slightly different input
parameters, EOS, and opacity tables used.

\section{Observational uncertainties}\label{sec:obs-unc}

\subsection{Cluster distance}\label{sec:distance}

As a next step we examine observational uncertainties
affecting our results and begin with the cluster distance. 
According to Table~4 in \citet{cop11}, previous distance
determinations for M5 fall in the range $14.32 \leq (m-M)_{0} \leq
14.67$. A weighted mean over the results compiled in their Table~4,
but including also their own distance determination, gives for the
cluster a distance modulus of $(m-M)_{0} = 14.43 \pm 0.02$~mag.
However, several of these distances are based on HB and RR Lyrae
stars, whose properties also depend on the presence of additional
cooling in their RGB progenitors. Therefore, to avoid circular
reasoning, we rely only on the distance determinations which do not
depend on evolved distance indicators which may be affected by the
cooling process which we want to constrain, as is the case with HB,
RR~Lyrae, and WD stars.

We thus adopted the distance modulus Eq.~(\ref{distance})
derived via main-sequence fitting by \citeauthor{L05} (\citeyear{L05}).
This result and corresponding error $0.11$ already take into account
interstellar extinction. This distance modulus
is in excellent agreement with the weighted mean value derived
above, on the basis of different distance indicators.

\begin{figure}[h]
   \centering
   \includegraphics[width=0.85\columnwidth]{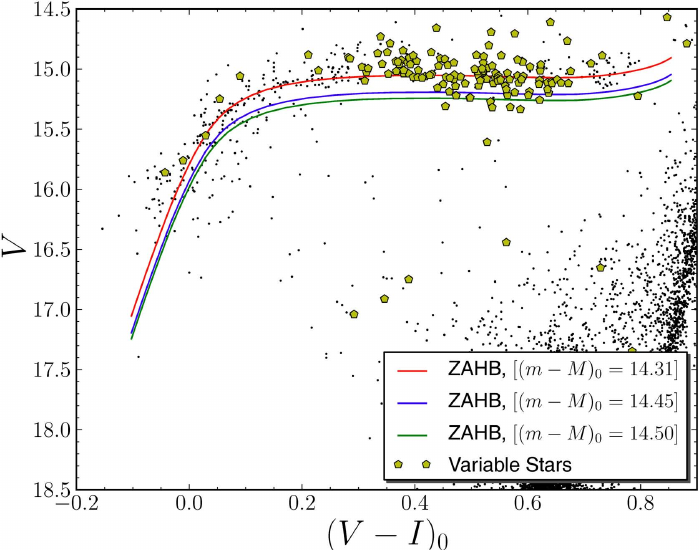}
   \caption{PGPUC ZAHB loci for different assumptions on the distance modulus
   overplotted on the HB region of M5. Variable stars from the
   \citet{clement} catalog are plotted in yellow.}
   \label{zahb}
\end{figure}

To check if the adopted distance modulus is consistent with our
model calculations, in Fig.~\ref{zahb} we overplot a ZAHB locus,
computed with PGPUC for the adopted chemical composition and age, on
the HB region of the M5 CMD, under different assumptions for the
distance modulus. We can see that the adopted distance modulus does
appear to provide a reasonable approximation to the lower envelope
of the observed HB distribution.

\subsection{Photometric calibration}\label{sec:calib}

\citeauthor{st2005} (\citeyear{st2005}) has revised the photometric
calibration of the standard fields that are used to calibrate our M5
photometry. In particular, this author performs a comparison with
\citeauthor{l92}'s (\citeyear{l92}) standard fields, finding that
the difference between Landolt's system and the adopted calibration
is not larger than $0.02$~mag in the $I$-band \citep[see Fig.~1
of][]{st2005}.

\subsection{Saturation}\label{sec:satur}

In \citet[][]{st2003}, the image-quality index $\chi$ is defined,
which shows us the level of agreement between the perceived
brightness profile for an object and the model PSF for the frame
where this object is measured. Reliable stars should have $\chi
\approx 1$, while significantly larger values of $\chi$ indicate
nonstellar objects or profiles corrupted by image defects.

This paper also defines the {\tt sharp} index, which is a
first-order estimate of the intrinsic angular radius of a resolved
source. For reliable stars, {\tt sharp} should be around zero.

\begin{figure}[h]
  \centering
  \includegraphics[width=0.8\columnwidth]{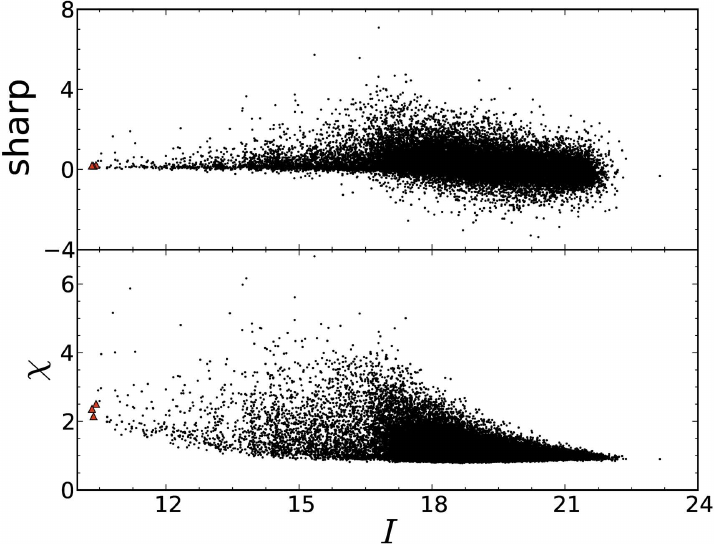}
  \caption{{\tt Sharp} (upper panel) and $\chi$ (lower panel).
   Red triangles: Three brightest stars.}
  \label{sharp}
\end{figure}

In Fig.~\ref{sharp} we plot {\tt sharp} and $\chi$ vs.\ the $I$-band
magnitude for our M5 photometry. The three brightest stars are shown
as red triangles. They are reliable based on these arguments. In
particular, if saturation were present we would have expected large
values for {\tt sharp} and $\chi$, but this is not seen in our data.

\begin{figure}[b]
  \centering
  \includegraphics[width=0.8\columnwidth]{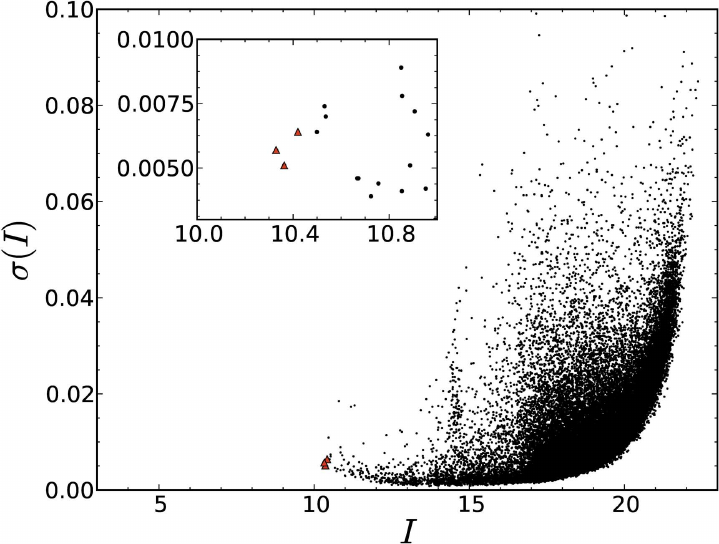}
  \caption{Photometric error $\sigma(I)$. Red triangles: three brightest
  stars. Inset: zoom around the brightest stars.
  The vertical plume of points at $I\approx 14.5$~mag consists
  of the RR Lyrae variables.}
  \label{errI}%
\end{figure}

If saturation were present, we might also expect large errors in the
derived $I$-band magnitudes. In Fig.~\ref{errI} we plot these errors
vs.\ the $I$ magnitudes themselves. Again the three brightest stars
are shown as red triangles. Their errors are all less than
$0.0064$~mag. Since the {\tt sharp}, $\chi$ and $\sigma(I)$ values
for the three brightest stars are all small, these stars appear not
to be saturated and can accordingly be safely used in our study.

On the other hand, we have also examined the individual $I$-band
magnitude measurements for each of these brightest stars, and found
that the rms deviation of one measurement is, in each case, around
0.05~mag. This is larger than the observational scatter for stars
that are a few magnitudes fainter, as is reflected in the increase
in $\sigma$ for magnitudes brighter than $I \approx 12$ in
Fig.~\ref{errI}. An increase in photometric uncertainties for the
very brightest stars is commonly seen in DAOPHOT reductions, and may
be due to incipient saturation:  when individual pixels in a star
image approach or exceed the known saturation level of the detector,
the software attempts to estimate the star's brightness by fitting
the model PSF only to unsaturated pixels in the flanks and wings of
the stellar profile.  Since these magnitude estimates are based on
fewer pixels, and on pixels that are less sensitive to brightness
than the central pixels of fainter stars, these brightness
determinations are noisier than those obtained for fainter stars. In
the case of the three brightest giants in M5, we find that the
measured magnitudes, while noisy, are {\it not\/} correlated with
telescope aperture or exposure time.  We infer that the DAOPHOT
software's treatment of incipient saturation---while subject to
greater noise---is not significantly biased in these particular
images. There are enough independent measurements of the $I$-band
magnitudes for each star that, despite the increased
observation-to-observation dispersion, the standard error of the
{\it mean\/} magnitude is still estimated to be well under 0.01~mag
in each case.

\begin{figure}[b]
   \centering
   \includegraphics[width=0.8\columnwidth]{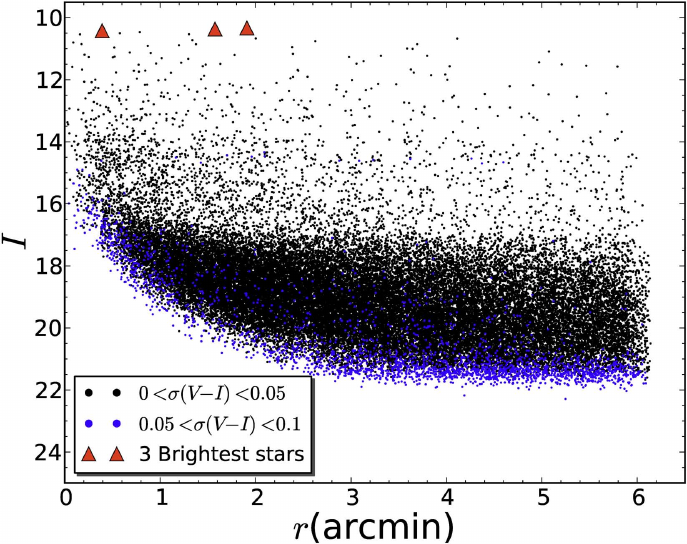}
   \caption{Radial distance (in arcminutes) vs.\ $I$-magnitude for different
   kinds of errors in $V-I$. Red triangles: Three brightest stars.}
   \label{crow}
\end{figure}

\subsection{Completeness and crowding}\label{sec:comp-crow}

Incompleteness is unlikely at the bright CMD levels that we are
considering in this study. However, to help us confirm that
incompleteness can be safely ignored, in Fig.~\ref{crow} we plot the
radial distance in arcminutes (between the center of M5 and each
star) versus the $I$-band magnitude for different ranges of
photometric errors. As an upper limit for photometric errors in
color we set $0.05<\sigma(V-I)<0.1$, since for higher errors the
photometry cannot be considered reliable. This range is plotted as
blue points. They can be taken as defining a reference locus beyond
which our photometric catalog can be considered incomplete.
Consistent with expectations, this reference locus is flat at large
distances from the cluster center, but moves towards brighter
magnitudes as one approaches the core. Stars much brighter than this
reference locus can be considered effectively complete at that
radius. Stars near the reference locus can be considered more than
normally unreliable and probably incomplete, whereas stars much
fainter than the reference locus are impossible to measure at that
radius. Our catalog is virtually 100\% complete down to the HB level
(where the RR Lyraes are) all the way to the center of the cluster.
Outside of $r=3$--4~arcmin, the catalog is virtually complete to
$I=21$~mag. As expected, the brightest stars in our sample, marked
as red triangles in Fig.~\ref{crow}, are much brighter than the
level at which incompleteness kicks in. We thus conclude that our
TRGB measurements are not affected by incompleteness.

On the other hand, and again as can be seen in Fig.~\ref{crow}, the
three brightest red giants in M5 fall quite close to the cluster
center \citep[recall that M5's half-light radius is 1.77~arcmin;][
2010 update]{wh96}, and can thus be plausibly affected by crowding.
To look further into this possibility, we have examined images from
the {\em Hubble Space Telescope}'s {\em Advanced Camera for Surveys}
(HST/ACS) archives. Looking at the three brightest stars in a
stacked HST/ACS Wide-Field Camera image, we see no stars that are
both within one or two arcseconds and within approximately five
magnitudes of the stars in question. Therefore, we do not expect
contamination in excess of one per cent or so, which is essentially
negligible for our purposes.

\subsection{Differential reddening}

\begin{figure}[b]
   \centering
   \includegraphics[width=0.9\columnwidth]{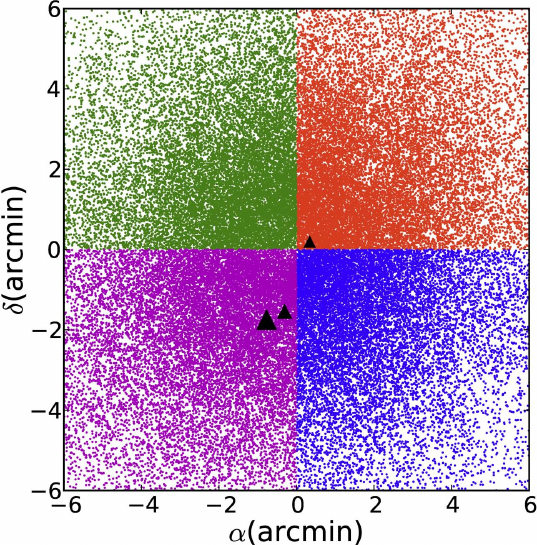}
   \caption{Differential reddening test. Here we show the four different
   quadrants of the cluster that were selected to produce the CMDs
   shown in Fig.~\ref{difred}. Black triangles denote the position of
   the three brightest cluster stars, with their sizes proportional to
   their apparent brightness.}
   \label{radec}
\end{figure}

\begin{figure}
   \centering
   \includegraphics[width=1\columnwidth]{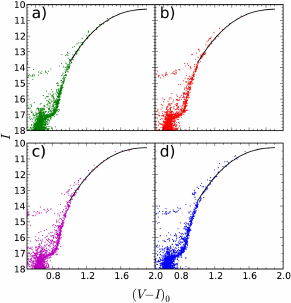}
   \caption{CMDs for the different quadrants shown in Fig.~\ref{radec} with
   the same color coding. The black line in all panels
   corresponds to the empirical RGB fit of Eq.~(\ref{eq:RGBfit}).}
   \label{difred}
\end{figure}

The amount of foreground reddening in the direction of M5 is quite
low, and a large amount of differential reddening that could change the relative
position of stars in the CMD and lead to wrong identifications of
RGB/AGB stars, is therefore unlikely.
To see if differential reddening is present in M5 at a
level that might affect our measurements, in Fig.~\ref{radec} we
divide the observed cluster field in four parts, producing separate
CMDs for each quadrant as shown in Fig.~\ref{difred}, color-coded
according to Fig.~\ref{radec}. Overplotting the CMD for all four
quadrants with our empirical RGB fit of Eq.~(\ref{eq:RGBfit}) in
Fig.~\ref{difred} reveals that the same fiducial line properly
describes all four quadrants. In addition, the giant branches are
narrow, consistent with the photometric errors. We thus conclude
that no significant amount of differential reddening is present in
the direction of the cluster.

To further verify this conclusion, we have examined the COBE/DIRBE
dust maps provided by \citet{sc1998}. Within a circle of 5~arcmin
radius centered on M5, the minimum and maximum values of $E(B-V)$
estimated from 100-micron dust emission are 0.0359 and 0.0398~mag,
respectively---again confirming the lack of significant differential
extinction across the face of the cluster.

\subsection{Summary}\label{sec:obs-sum}

 The observational TRGB brightness is that of the brightest
star $I_1$ given in Table~\ref{table:0}, shifted by the estimated
average brightness difference between TRGB and brightest star
$\langle\Delta_{\rm tip}\rangle$ of Eq.~(\ref{eq:statisticalshift}),
and corrected with the distance modulus of Eq.~(\ref{distance}),
i.e., $M_{I,\rm TRGB}^{\rm obs}=I_1-\langle \Delta_{\rm tip}\rangle
-(m-M)_0$. The error of $I_1$ is $\sigma_I=0.023$, which is
$\pm0.0057$ mag added in quadrature with the error of the
calibration of the photometry $\pm 0.02$~mag
(Sect.~\ref{sec:calib}). Saturation, completeness, and crowding
combined likely contribute less than $\pm 0.01$~mag to $\sigma_I$
(Sects.~\ref{sec:satur} and \ref{sec:comp-crow}). The statistical
error of the estimation of the TRGB relative to $I_1$ is
$\sigma_{\Delta_{\rm tip}}=0.058$. The largest error, $\sigma_{m-M}=0.11$, is
that of the distance modulus. Overall we thus find
\begin{equation}
M_{I,\rm TRGB}^{\rm obs}=-4.17\pm 0.13~{\rm mag}\,,
 \label{eq:photoerror}
\end{equation}
where we have added the errors in quadrature. Note that this is
appropriate even though the quantity $\Delta_{\rm tip}$ has a very
asymmetric distribution because, after convolution with the broad
Gaussian distance error distribution, we again have essentially a
Gaussian with a shifted mean and variance obtained by adding the
variances of the convolved distributions.

\section{Theoretical uncertainties}\label{sec:teo-unc}

We next address uncertainties that affect the predicted TRGB
$I$-band absolute magnitude $M_{I,{\rm TRGB}}$. Our benchmark model
parameters were described in Sect.~\ref{sec:pgpuc}. We use the
bolometric correction (BC) of \citet{W11} which cover the range
$\log g < 0$, which is not the case for most other BC prescriptions.
Moreover, \citet{W11} provide error estimates that we can use in our
assessment and, within this error estimate, agree with other BC
prescriptions.

\begin{figure*}[!t]
   \centering
   \begin{tabular}{cc}
   \includegraphics[width=0.77\columnwidth]{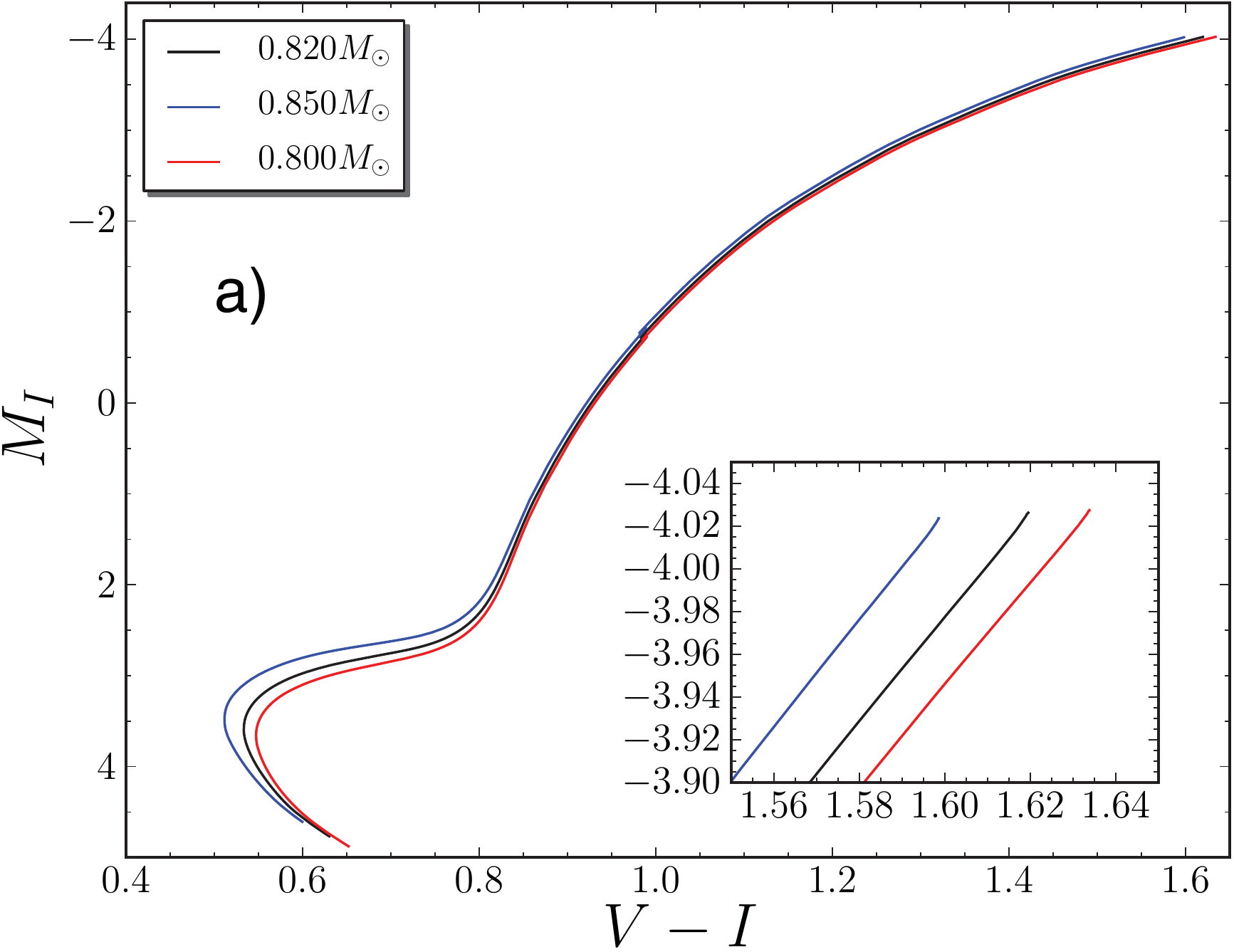} &
   \includegraphics[width=0.77\columnwidth]{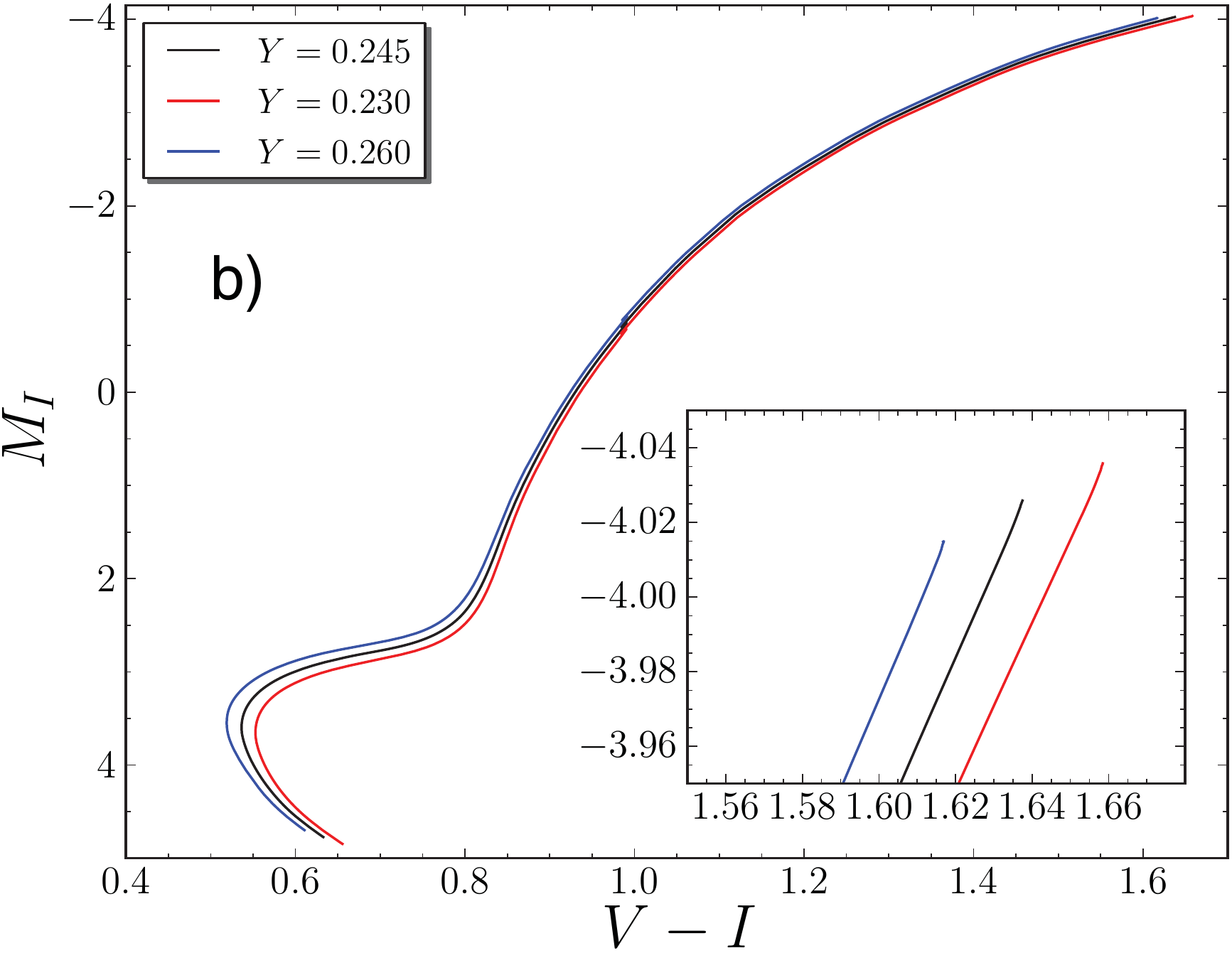} \\[10pt]
   \includegraphics[width=0.77\columnwidth]{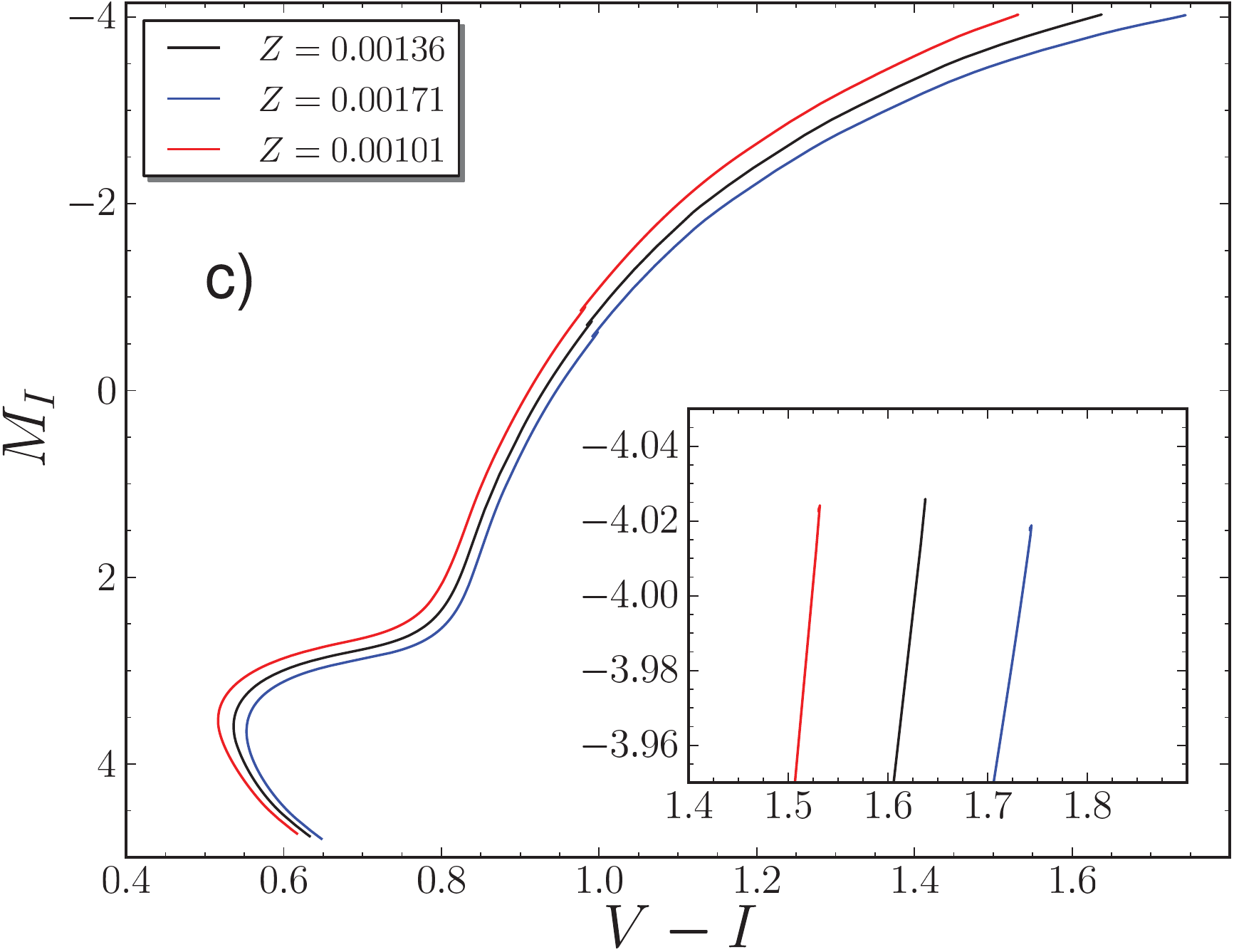} &
   \includegraphics[width=0.77\columnwidth]{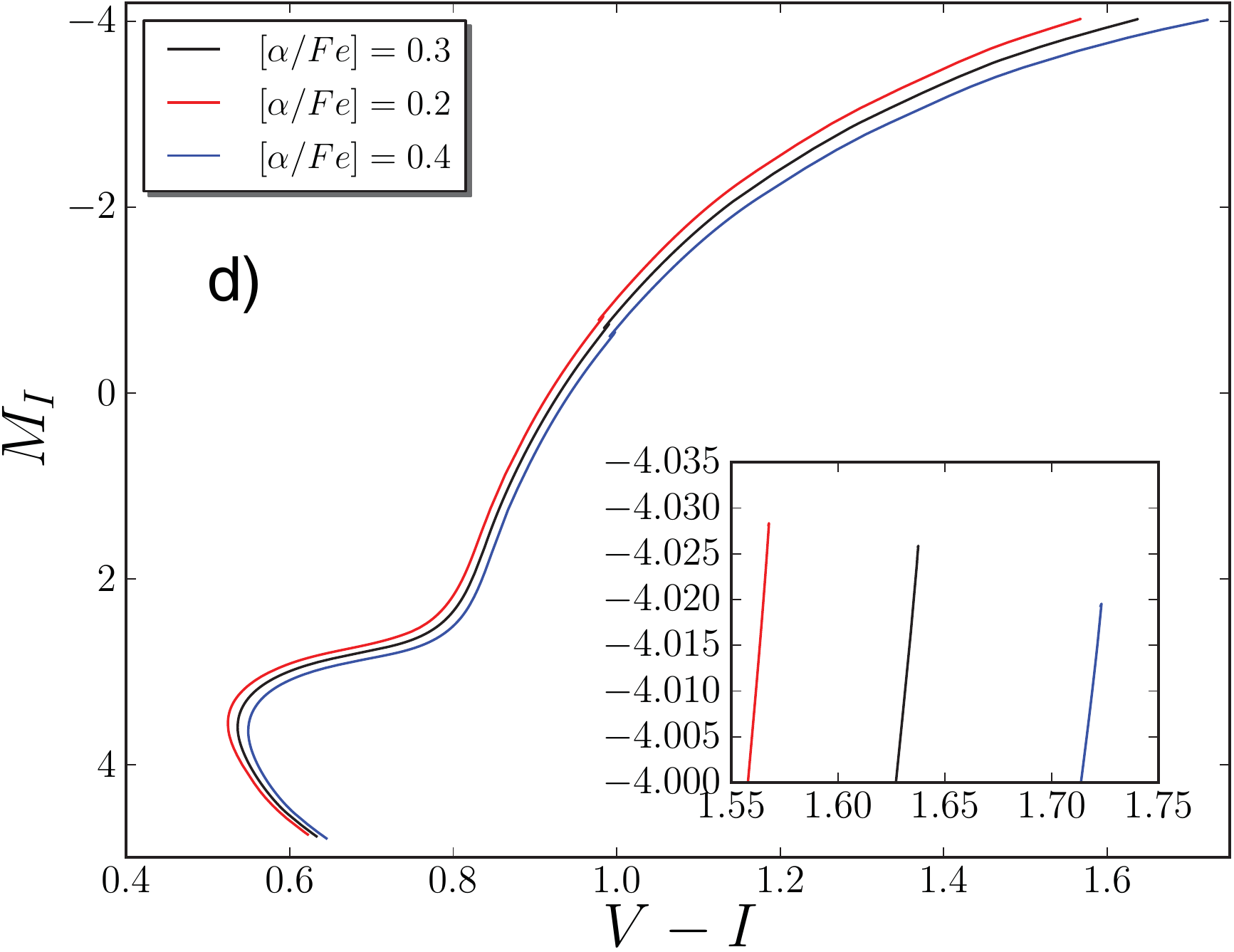}
   \end{tabular}
   \caption{Evolutionary tracks in the $M_I$, $V-I$ plane are shown for
   different: a)~Mass values, b)~$Y$ values, c)~$Z$ values,
   d)~$[\alpha/{\rm Fe}]$ ratios. The insets at the lower-right corner
   of each panel show zoom-ins around the TRGB position (note the
   different scales used in the different insets).}
   \label{comp1}
\end{figure*}

\subsection{Selected mass}

As mentioned in Sect.~\ref{sec:pgpuc}, we have selected an RGB mass
of $0.82 \, M_{\odot}$, as appropriate for the adopted chemical
composition and an age of about 13.8~Gyr. Uncertainties in GC ages
remain significant. For instance, \citet{adea10} estimate for M5, on
the basis of HST/ACS images, an age of $12.25\pm 0.75$~Gyr. However,
changing the age by $\pm 1$~Gyr changes the mass of the star at the
TRGB by only about $\pm 0.015 \, M_{\odot}$. Here we evaluate the
level of uncertainty in $M_{I,{\rm TRGB}}$ brought about by a
generous uncertainty in age of $1.5$~Gyr. Selecting masses of $0.80
\, M_{\odot}$ and $0.85\,M_{\odot}$ (see Fig.~\ref{comp1}a), we
obtain
\begin{eqnarray}
M_{I,{\rm TRGB}}^{0.85\,M_{\odot}} -M_{I,{\rm TRGB}}^{0.82\,M_{\odot}}&=&0.003~{\rm mag}\,,
\nonumber\\
M_{I,{\rm TRGB}}^{0.82\,M_{\odot}} - M_{I,{\rm TRGB}}^{0.80\,M_{\odot}}&=&0.001~{\rm mag}.
\nonumber
\end{eqnarray}
We conclude that uncertainties in the age, and thus in the RGB mass,
affect $M_{I,{\rm TRGB}}$ at the level of about $\pm 0.002$~mag
only.

\subsection{Helium abundance}

The He abundance $Y$ is uncertain for several reasons, including
statistical and systematic errors in the primordial He abundance
$Y_{\rm p}$ \citep[e.g.,][and references therein]{eaea10} and the He
enrichment parameter $\Delta Y/\Delta Z$ \citep[e.g.,][and
references therein]{pgpuc}. The primordial He abundance is most
strongly constrained by studies of extragalactic H\,{\sc ii}
regions. Based on several recent such studies
\citep[e.g.,][]{yiea07,it10,eaea10}, we estimate a range of
uncertainty in the primordial He abundance of $0.230 <Y_{\rm p} <
0.260$. At the relevant metallicity $Z$, the error arising from
uncertainties in $\Delta Y / \Delta Z$ is negligible in comparison.
The impact of variations in $Y$ at this level upon the TRGB position
is shown in Fig.~\ref{comp1}b. We conclude that $M_{I,{\rm TRGB}}$
changes by about $\pm 0.010$~mag for a change in $Y$ by $\pm 0.015$.

\subsection{Iron abundance}

As explained in Sect.~\ref{sec:pgpuc}, we have adopted an iron
abundance with respect to the Sun ${\rm [Fe/H]}=-1.33 \pm 0.02$~dex,
from \citet[]{Carr09}. The level of uncertainty in the cluster's
[Fe/H] value was critically evaluated by \citet{kmcw10}, who
recommended a value ${\rm [Fe/H]}=-1.33 \pm 0.02 ({\rm stat.}) \pm
0.03 ({\rm syst.})$~dex. Indeed, the increased error bar appears to
better accomodate other recent [Fe/H] determinations, including
\citet{dlea11} and \citet{isea12}. Assuming $[\alpha/{\rm Fe}] =
+0.3$ (see Sect.~\ref{sec:alphaFe}), the error associated to the
iron abundance $\sigma_{\rm [Fe/H]}=\pm 0.036$~dex translates into
an error in the global metallicity of $\sigma_Z=\pm 0.00011$.
However, the uncertainty in the solar metallicity proper
\citep[see][for a review and references]{mc12} also adds up to the
budget, leading to a final uncertainty estimate in $Z$ of $\pm
0.00035$. The impact of variations in $Z$ at this level upon the
TRGB position is shown in Fig.~\ref{comp1}c. This leads to a change
in $M_{I,{\rm TRGB}}$ of $+0.070$~mag, which, perhaps somewhat
surprisingly, is positive for both the higher and lower $Z$ values,
possibly due to the substantial curvature of the TRGB in the CMD.

\subsection{$[\alpha /{\rm Fe}]$ ratio}\label{sec:alphaFe}

As explained in Sect.~\ref{sec:pgpuc}, we have adopted an
$\alpha$-element-to-iron enhancement ratio of $[\alpha/{\rm Fe}] =
+0.3$ (for fixed [Fe/H] this therefore increases the value of $Z$ of the
models), based on \citet{dlea11}. Based on this latter study, one can
estimate a reasonable uncertainty in this ratio of about $\pm 0.1$,
ignoring the oxygen determinations during to their large associated
errors. Figure~\ref{comp1}d shows a comparison between evolutionary
tracks with different $[\alpha/{\rm Fe}]$ values covering this
range. An increment in $[\alpha/{\rm Fe}]$ implies a fainter TRGB,
as found by \citet{cassisi04}. We finally estimate a $1\sigma$ error
in $M_{I,{\rm TRGB}}$ of $\mp 0.0044$~mag.

\begin{figure*}[!t]
   \centering
   \begin{tabular}{cc}
   \includegraphics[width=0.88\columnwidth]{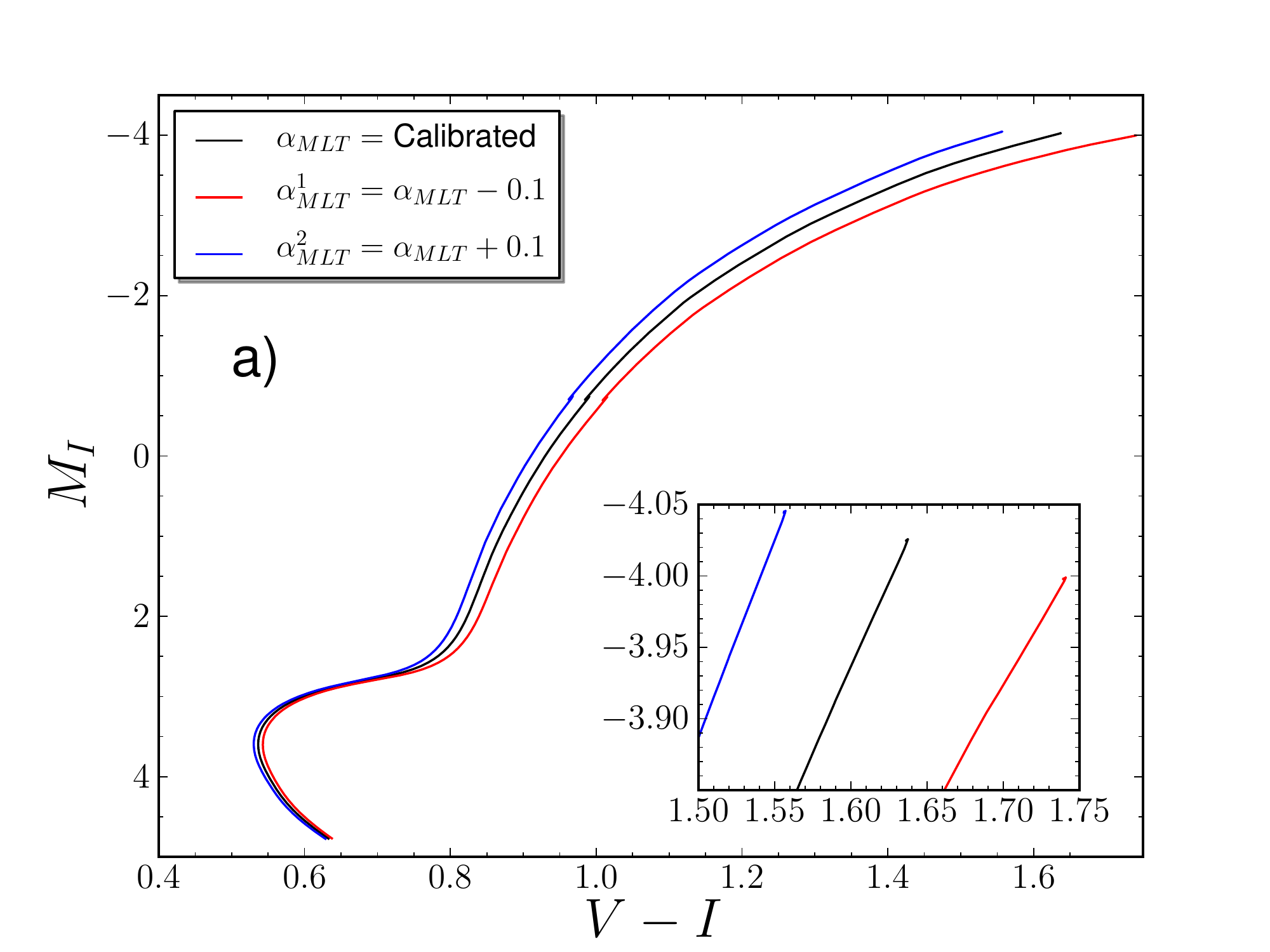} &
   \includegraphics[width=0.88\columnwidth]{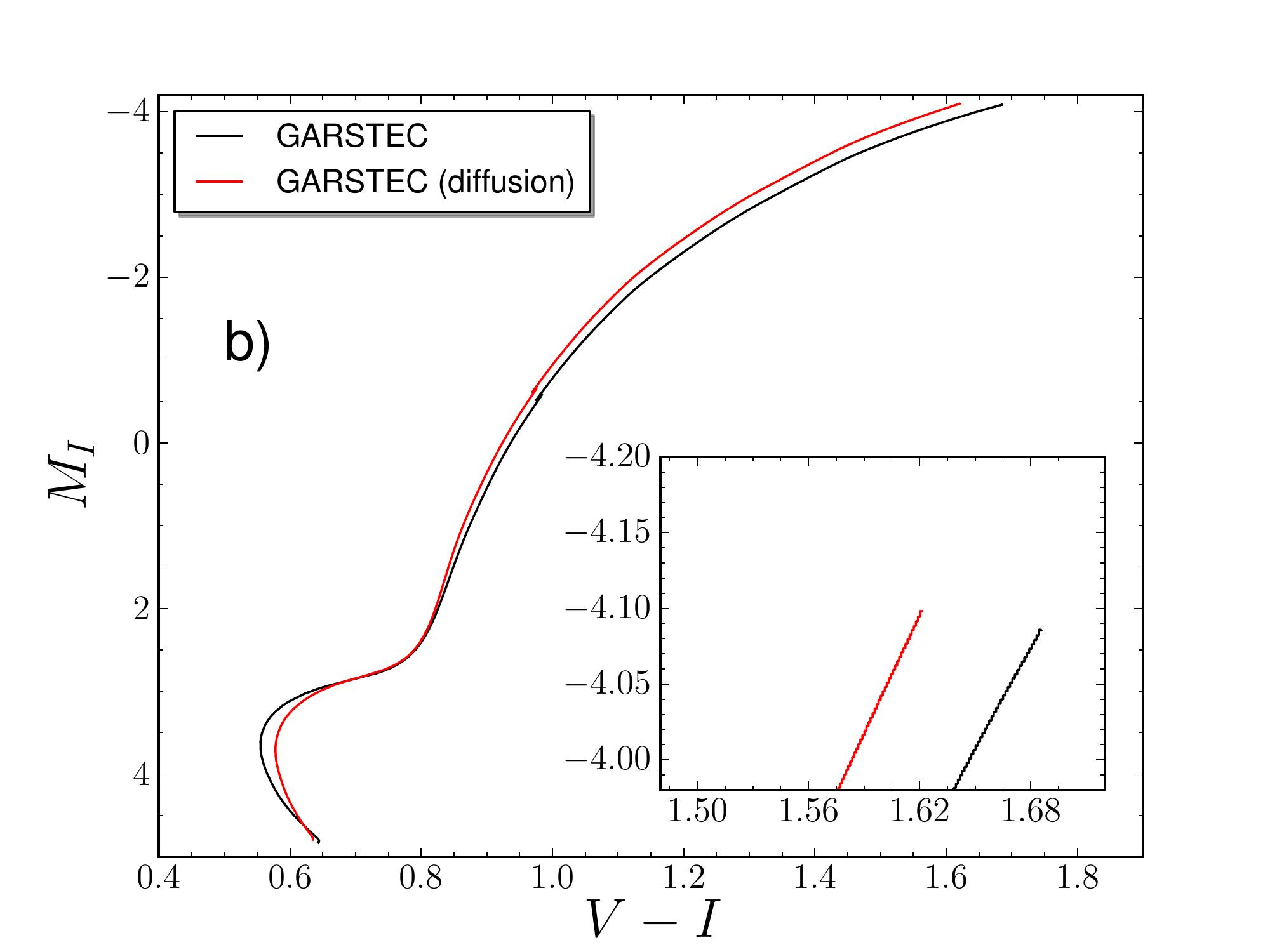}\\[10pt]
   \includegraphics[width=0.77\columnwidth]{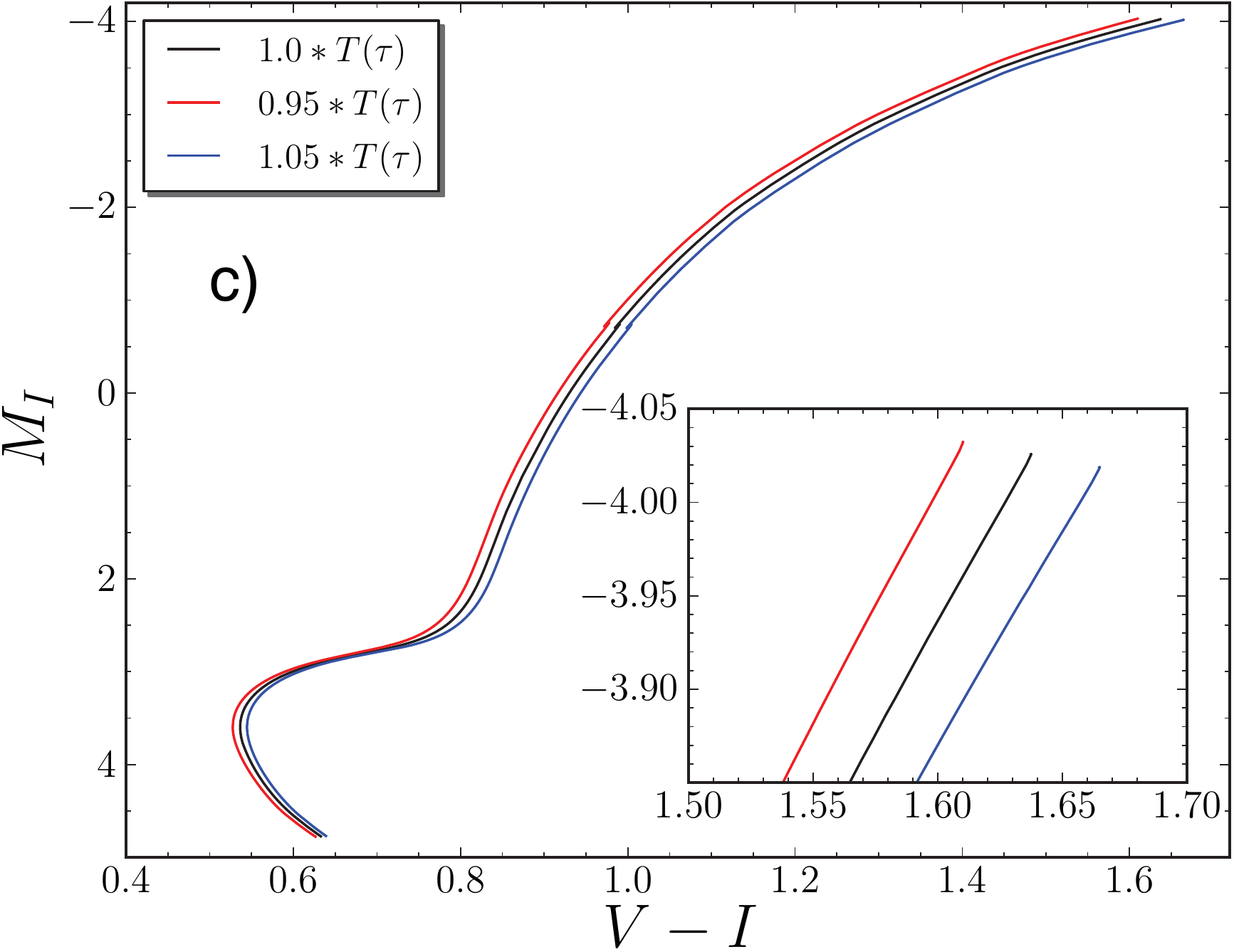} &
   \includegraphics[width=0.77\columnwidth]{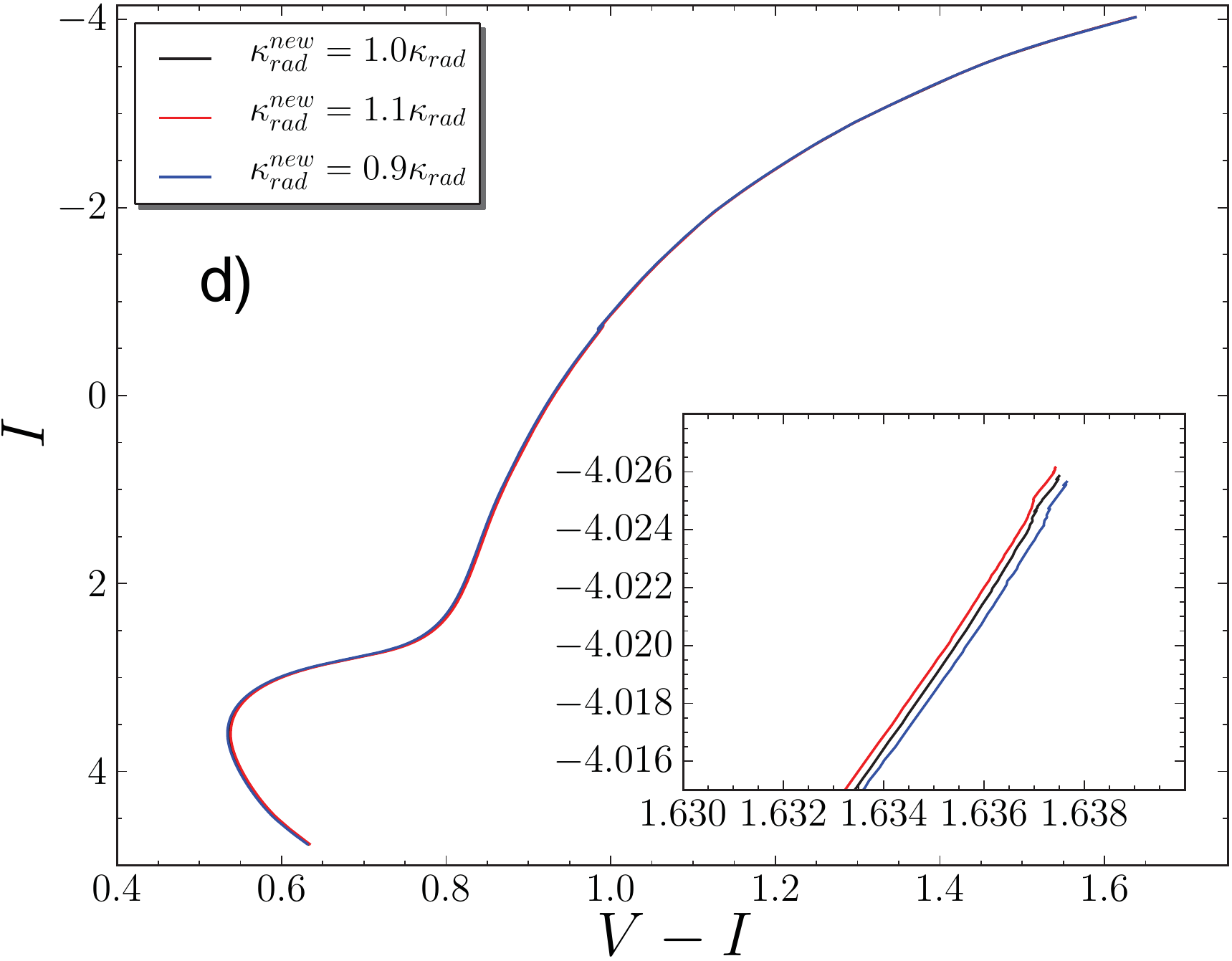}
   \end{tabular}
   \caption{As in Fig.~\ref{comp1}, but for different: a)~$\alpha_{\rm MLT}$
   values, b)~treatment of diffusion, c)~boundary conditions,
   d)~$\kappa_{r}$ values.}
   \label{comp2}%
\end{figure*}

\subsection{Mixing-length parameter, overshooting and
diffusion}\label{sec:mixing}

In spite of its limitations \citep[e.g.,][]{lk12}, we have adopted
the mixing-length theory (MLT) as described in \citet{pgpuc}. In
other words, following the usual procedure, the mixing-length
parameter $\alpha_{\rm MLT}$ of convection theory has been fixed to
reproduce the present-day solar parameters, allowing for a quite
satisfactory match of the CMDs of GCs over a wide range of
metallicities \citep[e.g.,][and references
therein]{fs99,rpea02,ffea06,dvea08,sc12}. From such studies, an
uncertainty of $\pm 0.1$ in $\alpha_{\rm MLT}$ (for a given
evolutionary code) appears reasonable \cite[see][]{S02}.

In addition, an uncertainty due the calibration of $\alpha_{\rm
MLT}$ is needed. In \citet{pgpuc}, $\alpha _{\rm MLT}$ was
calibrated without atomic diffusion, but it is known that the
sound-speed profile of the Sun agrees better with helioseismological
data if one includes atomic diffusion \citep[see Sect. 5
of][]{basti04}. The difference between $\alpha _{\rm MLT}$
calibrated with and without diffusion is approximately $0.1$ from
the GARSTEC stellar evolution code \citep{garstec} and S.~Cassisi
(priv.\ comm.).

We finally adopt an uncertainty of $\pm 0.2$. Note that such changes
in $\alpha_{\rm MLT}$ also affect the amount of He that is dredged
up as a consequence of the first dredge up episode. The tracks
computed for different $\alpha_{\rm MLT}$ values (Fig.~\ref{comp2}a)
reveal that an uncertainty of $\pm 0.2$ in $\alpha_{\rm MLT}$
changes $M_{I,{\rm TRGB}}$ by $\mp 0.056$~mag.

According to \citet{lk12}, overshooting needs to be included to
bring MLT models into agreement with 3D simulations of the outer
layers of RGB stars. In this sense, overshooting had previously also
been suggested as a means to reconcile predicted and observed
positions of the RGB luminosity function ``bump'' \citep[see][for a
review and references to early work and also \citeauthor{sc12}
\citeyear{sc12} for a more recent review]{S02}. Typically,
overshooting is modeled, as an extension of the MLT formalism, by
extending the formal boundary of the convective zone by a fraction
\citep[typically around 20--25\%; e.g.,][and references
therein]{dvdbea06,scea11} of the local pressure scale height. In
this way, it can be anticipated that uncertainties in the treatment
of overshooting will play a minor role, compared with the
uncertainty in $\alpha_{\rm MLT}$ proper, as far as the TRGB
position goes. We accordingly ignore overshooting as a separate
source of uncertainty.

Atomic diffusion can also be a source of uncertainty for $M_{I,{\rm
TRGB}}$. Compared with the case without diffusion, evolutionary
tracks of low-mass stars which take into account fully efficient
atomic diffusion include the following: lower envelope He abundances
(by $\sim 0.01$, after the main sequence); changes in the detailed
stellar evolution path close to the main-sequence turnoff point
\citep[][]{S02}; and an increment in the He core mass (by $\sim
0.004 \, M_{\odot}$, at the TRGB). These effects cannot be studied
separately. Unfortunately, PGPUC does not yet include atomic
diffusion, and so it is not possible for us to study this question.
Therefore, we have obtained evolutionary tracks with and without
diffusion from the GARSTEC stellar evolution code \citep{garstec} using
our stellar parameters. These tracks
(Fig.~\ref{comp2}b) imply that diffusion changes $M_{I,{\rm TRGB}}$
by $-0.0123$~mag. Since diffusion is partly compensated by
turbulence, which is not included in the GARSTEC calculations, the
likely level of change brought about by the inclusion of diffusion
is around $-0.0062$~mag in $M_{I,{\rm TRGB}}$.

\begin{figure*}
   \centering
   \begin{tabular}{cc}
   \includegraphics[width=0.77\columnwidth]{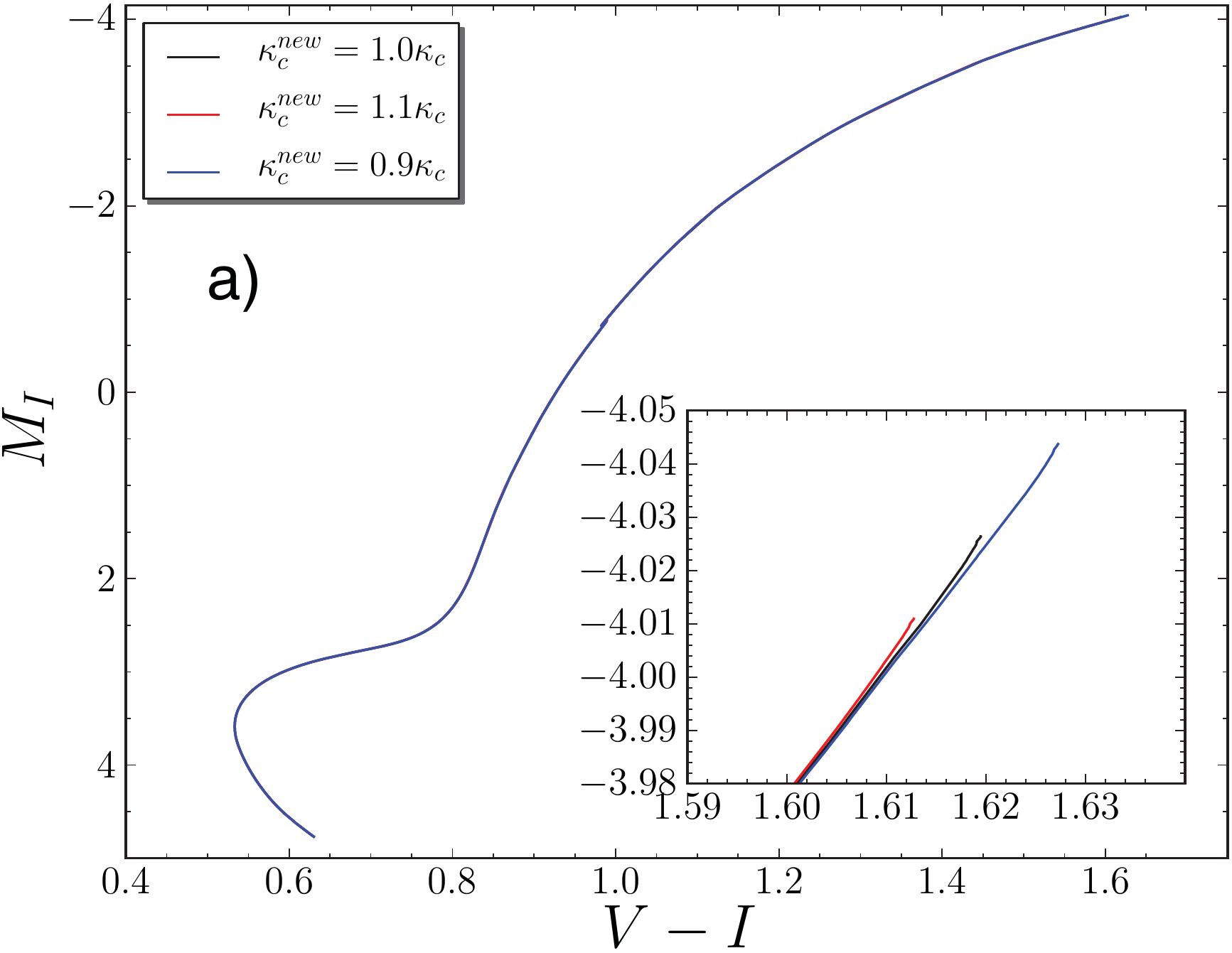} &
   \includegraphics[width=0.77\columnwidth]{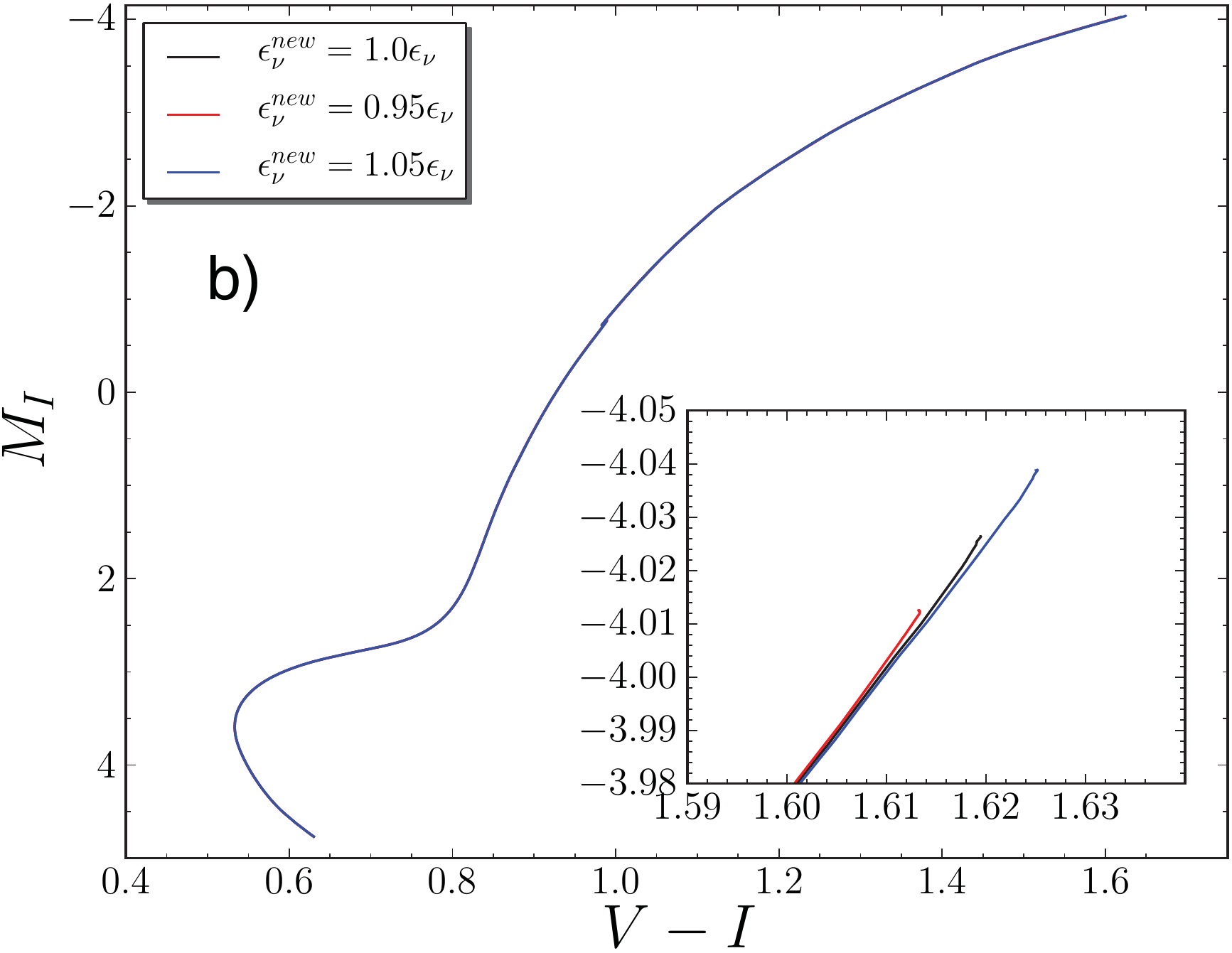} \\[10pt]
   \includegraphics[width=0.77\columnwidth]{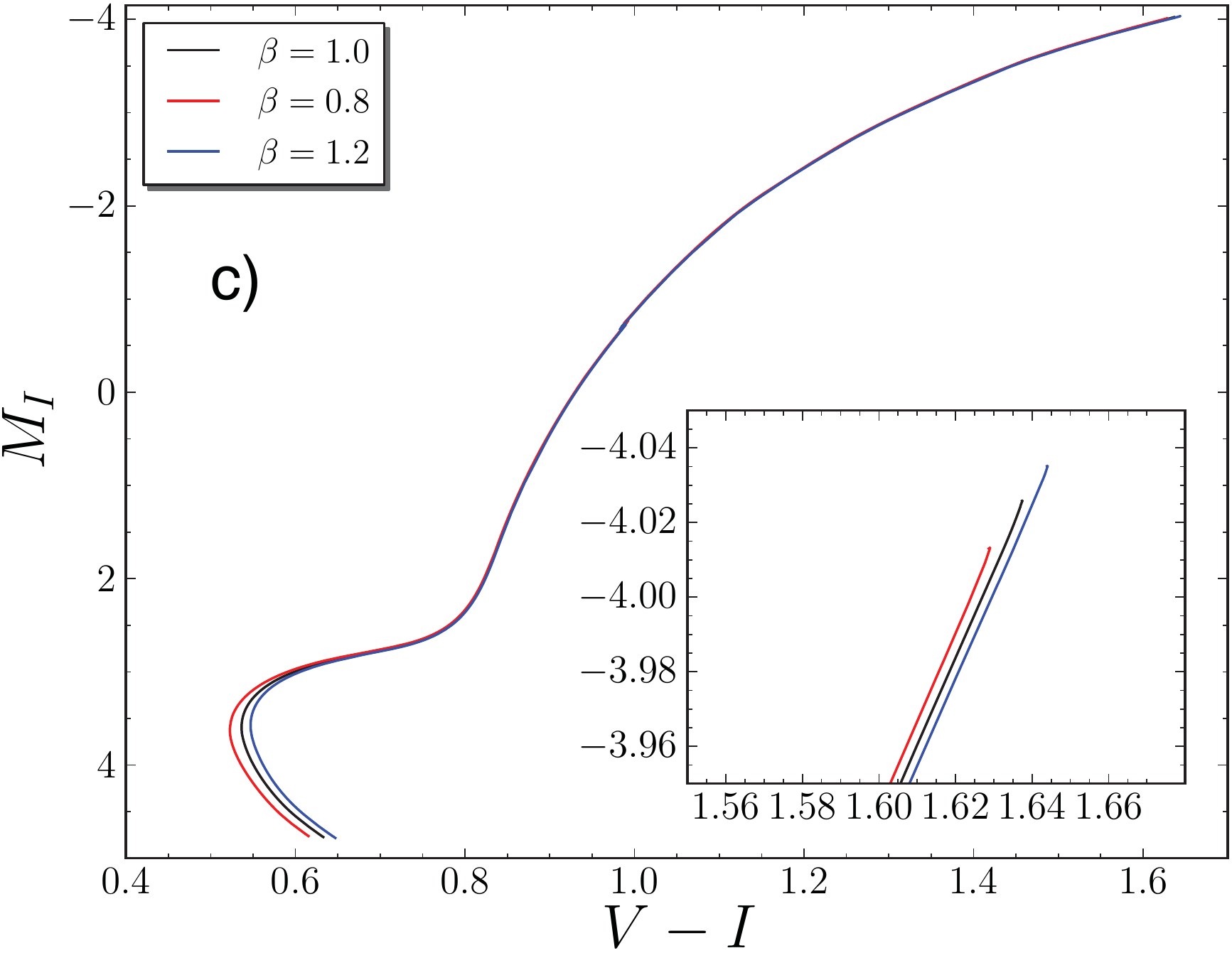} &
   \includegraphics[width=0.89\columnwidth]{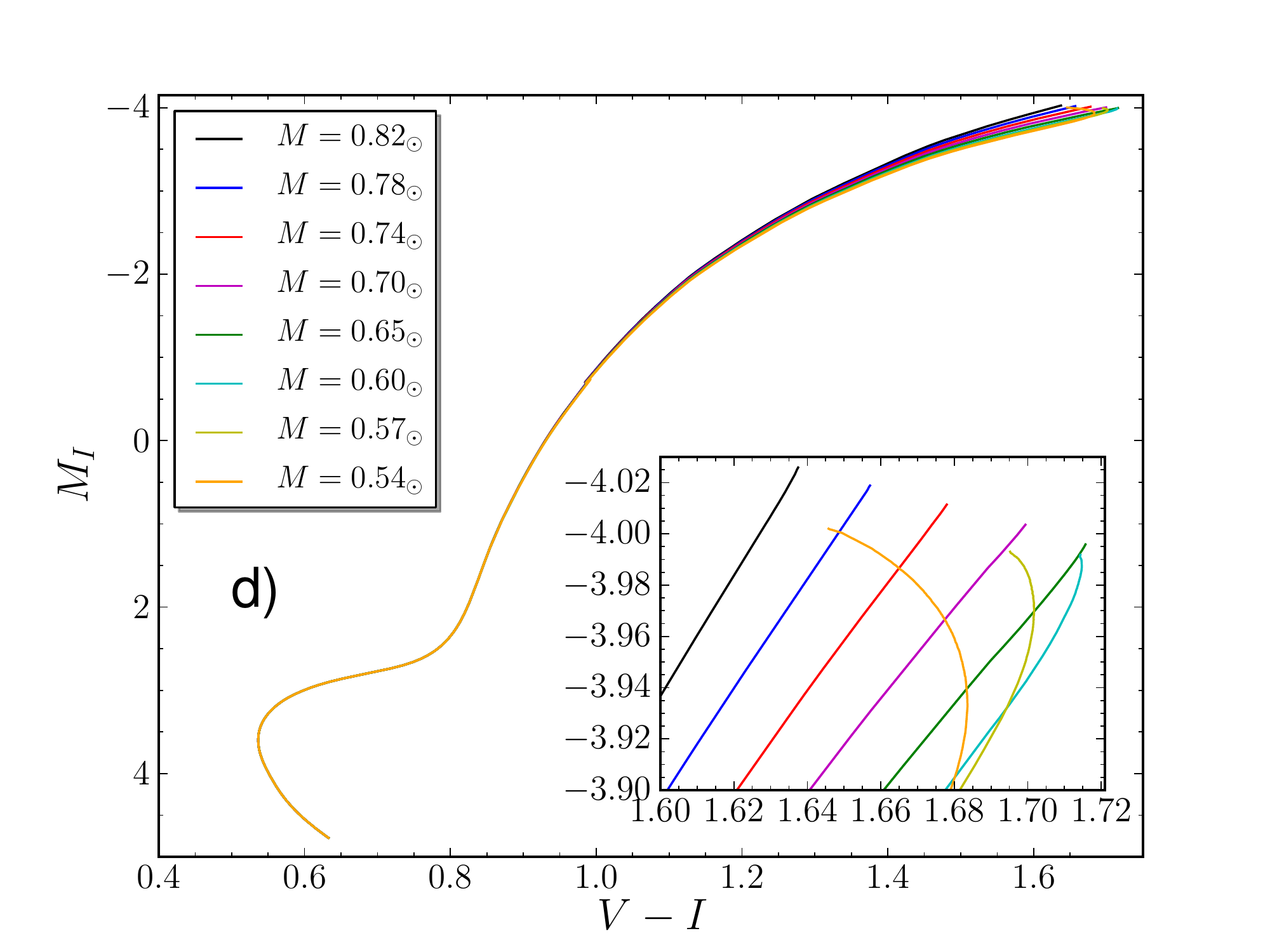}
   \end{tabular}
   \caption{As in Fig.~\ref{comp1}, but for different: a)~$\kappa_{c}$ values,
   b)~neutrino emissivities, c)~screening factors, d)~mass loss
   parameters.}
   \label{comp3}
\end{figure*}

\subsection{Boundary conditions}

Different treatments of surface boundary conditions will naturally
change the location of the TRGB (see, e.g., \citeauthor{S02}
\citeyear{S02}; \citeauthor{dvea08} \citeyear{dvea08}). In this
sense, PGPUC uses the $T(\tau)$ relation given in Eq.~(1) of
\citet{pgpuc}. Figure~7 of \citet{tanner2012} shows that different
prescriptions for the treatment of the outer regions of a red giant
may change the resulting $T(\tau)$ relation by approximately
$\pm5\%$. In Fig.~\ref{comp2}c we show the impact of such a change,
after recalibrating the solar model accordingly. We conclude that
such an uncertainty leads to an $M_{I,{\rm TRGB}}$ uncertainty of
$\mp 0.0068$~mag.

\subsection{Radiative opacities}

Under non-degenerate conditions, radiative transfer is the prime
mechanism of energy transfer. But as the luminosity of the TRGB is
regulated mainly by the degenerate He core, one expects that
uncertainties in the radiative opacities $\kappa_{\rm rad}$ are not
of primary importance for the evolution close to the TRGB. The
current best estimates of uncertainties in $\kappa_{\rm rad}$ tend
to fall around the 5\% level, but there have been suggestions,
raised primarily in the framework of the so-called ``solar abundance
problem,'' that the uncertainties in $\kappa_{\rm rad}$ may be
higher, possibly reaching up to 20--30\% at some points in the
stellar interior (see \citeauthor{mc12} \citeyear{mc12} for a recent
review and references).

To evaluate the impact of $\kappa_{\rm rad}$ uncertainties, we have
changed $\kappa_{\rm rad}$ by $\pm 10\%$ and recalibrated the solar
model accordingly, resulting in a $M_{I,{\rm TRGB}}$ change by only
$\mp 0.00025$~mag (Fig.~\ref{comp2}d). As suspected, uncertainties
in $\kappa_{\rm rad}$ are negligible for the predicted TRGB
position.

\subsection{Conductive opacities}

Energy transport, in the degenerate conditions characterizing the He
cores of RGB stars, occurs primarily by electron conduction.
Accordingly, the conductive opacity $\kappa_{\rm cond}$ plays a
crucial role in the determination of the He core mass at the TRGB
\citep[e.g., CFP96;][]{scea07,mc09}. It is difficult to evaluate the
current level of uncertainty in $\kappa_{\rm cond}$, because this
uncertainty is expected to be both temperature- and
density-dependent. Recently, \citet{gvea12} adopted a level of 5\%
uncertainty as a reasonable average uncertainty in their RGB models.
According to A.~Potekhin (2012, priv.\ comm.), an uncertainty of
$\pm 10\%$ should provide a reasonable estimate of the remaining
uncertainties in this key physical ingredient. We have accordingly
computed additional RGB tracks in which we changed $\kappa_{\rm
cond}$ by $\pm 10\%$, again recalibrating the solar model
accordingly, as shown in Fig.~\ref{comp3}a. The corresponding
uncertainty in $M_{I,{\rm TRGB}}$ is $\pm 0.016$~mag.

\subsection{Nuclear reaction rates}

Uncertainties in the main thermonuclear reactions rates considered
in PGPUC may affect the TRGB position. The main reactions of
hydrogen burning via the proton-proton chain are ${}^{3}\textrm{He}
+ {}^{3}\textrm{He} \rightarrow {}^{4}\textrm{He}+2p$,
${}^{7}\textrm{Be} + e^{-} \rightarrow {}^{7}\textrm{Li}+\nu _{e}$
and ${}^{7}\textrm{Be} + {}^{1}\textrm{H} \rightarrow
{}^{8}\textrm{Be}+\gamma$. For the CNO cycle, the reaction
considered is $^{14}{\rm N}(p,\gamma)^{15}{\rm O}$. It is the
slowest reaction in the CN cycle, sets the pace for the operation of
the cycle in equilibrium, and hence is the most important one for
our purposes. For helium burning, the reactions considered are
$^{4}{\rm He}(\alpha \alpha, \gamma)^{12}{\rm C}$ and $^{12}{\rm
C}(\alpha, \gamma) ^{16}{\rm O}$. Each of these reactions are
assumed to operate in chemical equilibrium.

In Table~\ref{table:1} we present the changes in $M_{I,{\rm TRGB}}$
that are brought about by uncertainties in the different nuclear
reaction rates, after recalibrating the solar model as needed in
each case. The latter uncertainty levels are adopted directly from
the original sources of the nuclear reaction rates used in PGPUC,
and are similar, for instance, to those adopted in \citet{gvea12},
and also of the same order of magnitude as estimated by
\citet{eaea11}. We note from Table~\ref{table:1} that  the
triple-$\alpha$ reaction is the one that impacts TRGB the most. In
summary, our calculations allow us to estimate a global uncertainty
in $M_{I,{\rm TRGB}}$ at the level of $\pm 0.019$~mag, as a
consequence of uncertainties in the relevant nuclear reaction rates.

\begin{table}
\caption{$M_{I,{\rm TRGB}}$ uncertainties caused by nuclear reaction
rates}
\label{table:1} \centering
\begin{tabular}{lll}
\hline\hline
\textbf{Nuclear reaction }  &  Range & Change in $M_{I,{\rm TRGB}}$ \\
\hline
${}^{1} \textrm{H} + {}^{1} \textrm{H} \rightarrow  {}^{2} \textrm{H} + e^{+}+\nu _{e} $     & $\pm 3\%$  &  $\pm 4.06 \times 10^{-4}$~mag \\
${}^{3}\textrm{He} + {}^{3}\textrm{He} \rightarrow {}^{4}\textrm{He}+2p$               & $\pm 2\%$ & $\pm 3.39 \times 10^{-4}$~mag \\
${}^{3}\textrm{He} + {}^{4}\textrm{He} \rightarrow {}^{7}\textrm{Be}+\gamma$           &  $\pm 6\%$ & $\pm 3.70 \times 10^{-4}$~mag \\
${}^{7}\textrm{Be} + e^{-} \rightarrow {}^{7}\textrm{Li}+\nu _{e}$ & $\pm 10\%$ & $\pm 2.27  \times 10^{-3}$~mag   \\
${}^{7}\textrm{Be} + {}^{1}\textrm{H} \rightarrow {}^{8}\textrm{Be}+\gamma$                     & $\pm 3\%$ & $\pm 2.03 \times 10^{-3}$~mag   \\
${}^{12}\textrm{C} + {}^{4}\textrm{He} \rightarrow {}^{16}\textrm{O}+\gamma $             & $\pm 10\%$ &  $\pm 1.25 \times 10^{-4}$~mag \\
${}^{4}\textrm{He} + {}^{4}\textrm{He} \rightarrow {}^{8}\textrm{Be}+\gamma$             & $\pm 19\%$ &  $\pm 1.39 \times 10^{-2}$~mag  \\
${}^{8}\textrm{Be} + {}^{4}\textrm{He} \rightarrow {}^{12}\textrm{C}+\gamma$             & $\pm 10\%$ &   $\pm 7.43 \times 10^{-3}$~mag  \\
${}^{14}\textrm{N} + p \rightarrow {}^{15}\textrm{O}+\gamma$             & $\pm 15\%$ &  $\mp 9.58 \times 10^{-3}$~mag  \\
\hline
{\bf TOTAL} & & $\pm 1.87 \times 10^{-2}$~mag \\
\hline
\end{tabular}
\end{table}

\subsection{Screening of the nuclear reactions}

In addition to the nuclear reaction rates proper, the corresponding
screening corrections must also be taken into account. These
corrections have been discussed for a long time, and have included
electrostatic screening \citep{sal54} and dynamical screening
\citep[e.g.,][]{shaviv00}. In this sense, while the recent
experimental results of \citet{faea10} are claimed to ``fully
corroborate'' the validity of the \citeauthor{sal54} results, the
results of recent dynamic screening calculations \citep[][and
references therein]{md11} provide a rather different perspective,
not being able to confirm the significant enhancements in the
nuclear reaction rates over their unscreened values that are
predicted in the static regime.

From a practical perspective, it is useful to note that
\citet{weiss01} have implemented different screening factors in a
solar model, using the \citet{gs98} abundances for the Sun,
concluding that a deviation of the screening factor by $\pm 10\%$ is
strongly discrepant with the solar sound speed profile estimated
with helioseismology. Recently, new and highly sophisticated
estimations of the solar abundances have become available
\citep[e.g.,][]{ags05,maea09,caff011}, but these new abundances
further increase the differences between theoretical estimations and
helioseismological measurements of the solar sound speed profile.
Several hypotheses have been suggested to solve this ``solar
abundance problem'' \citep[see, e.g.,][for a recent review and
references]{mc12}. In \citet{weiss08}, it was pointed out that an
increase of the screening factor by 10--15\% over the static values
can recover the agreement with the seismic sound speed profile in
the envelope, when the \citet{ags05} chemical composition is used.
However, in the center of the Sun such an increase in the screening
factor makes the sound speed profile deviate significantly from the
helioseismological values \citep[see also][]{mgea11}. Taking these
considerations into account, here we conservatively adopt an
uncertainty of $\pm 20\%$ in the screening factors. Tracks computed
with an enhancement factor $\beta$ in the range 80--120\% over the
reference (static) case are shown in Fig.~\ref{comp3}c, where the
corresponding solar models were also recalibrated as required. The
associated uncertainty in $M_{I,{\rm TRGB}}$ is $\pm 0.011 $~mag.

\subsection{Neutrino emission}

The neutrino emission rate used in PGPUC is described in
\citet{H94}, where an analytic expression for the plasma neutrino
emission rates is provided (based mosltly on the work of \citet{bra93}), which is likely correct at the $\pm 5\%$
level \citep[see][for a recent discussion and additional
references]{gvea12}. On the upper RGB, neutrino emission is the most
important energy-loss mechanism from the center of the star.
Therefore, any change in this formulation will have a direct impact
on the predicted level of the TRGB. Changing the neutrino emission
rates by $\pm 5\%$, we find that $M_{I,{\rm TRGB}}$ is changed by
$\mp 0.013$~mag (Fig.~\ref{comp3}~b).

\subsection{Mass loss}

It is well known \citep[e.g.,][]{cc93,ndcea96,nsea01} that the
position of the TRGB will change depending on the amount of mass
loss, with stars ``peeling off'' from the RGB locus at a lower
luminosity with increasing mass loss (see also Fig.~\ref{comp3}~d).
However, the TRGB position is expected to be dominated by RGB stars
that undergo the {\em least} amount of mass loss, since these are
expected to be brighter than their mass-losing counterparts. This is
why we originally adopted as reference case the evolutionary tracks
computed without mass loss. At any point in the life of the cluster,
however, and due to small-number statistics, the stars closest to the
TRGB may actually be the more strongly mass-losing giants that may be
present in the cluster.
\begin{figure}[h]
  \centering
   \includegraphics[width=0.8\columnwidth]{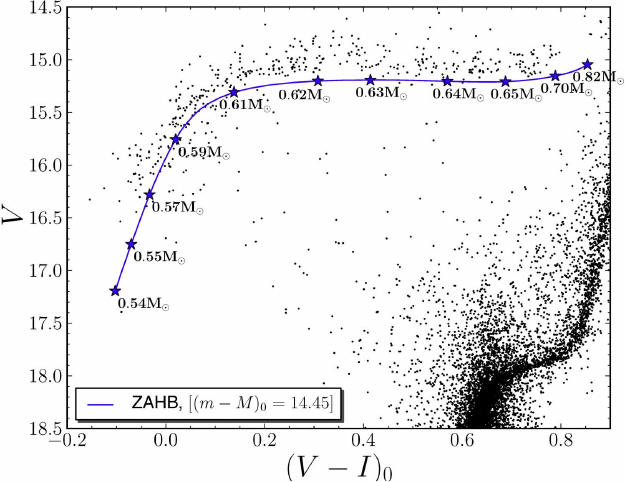}
   \caption{Comparison between the observed M5 CMD locus and PGPUC ZAHB
   predictions. The blue star symbols indicate the approximate
   locations of ZAHB stars with the indicated mass values,
   corresponding to overall RGB mass loss amounts of 0.19--$0.28 \, M_{\odot}$.}
   \label{loss}%
\end{figure}
Therefore, to evaluate the impact of mass loss, we need to constrain
the range in total mass loss that may be present among M5's RGB
stars. Judging from Fig.~\ref{loss}, where we overplot a zero-age HB
(ZAHB) locus obtained for the adopted chemical composition and age
with the observed CMD, the minimum masses reached by this cluster's
HB stars are around $0.54\, M_{\odot}$, which implies an upper limit
for the overall RGB mass loss of about $0.82-0.54 = 0.28 \,
M_{\odot}$, compared with an average value of order $0.82 - 0.63 =
0.19 \, M_{\odot}$ (a mass of $0.63 \, M_{\odot}$ corresponding
roughly to the mid point of the HB, as seen in Fig.~\ref{loss}) and
a minimum amount of mass loss of order $0.82 - 0.70 = 0.12 \,
M_{\odot}$, corresponding to the reddest stars along the ZAHB.
Taking these constraints into account, and using the PGPUC code with
the \citet{sc05,sc07} mass loss formalism, we find that for 
the minimum observed mass loss of $0.12\,M_\odot$ the value of
$M_{I,{\rm TRGB}}$ is shifted by $+0.022$~mag, whereas for the
maximum of $0.28\,M_\odot$ it is shifted by
$+0.024$~mag. 
However, $M_{I,{\rm TRGB}}$ does not depend linearly on mass loss, 
presenting a maximum at $0.22 \, M_{\odot}$ where 
$M_{I,{\rm TRGB}}=0.035$ mag (see Fig.~\ref{comp3}~d). 
We thus conclude that the uncertainty due to mass loss falls in the 
range $0.022$--$0.035$~mag.

\subsection{Equation of state (EOS)}
\label{sec:eos}

\begin{figure}[h]
   \centering
   \includegraphics[width=0.8\columnwidth]{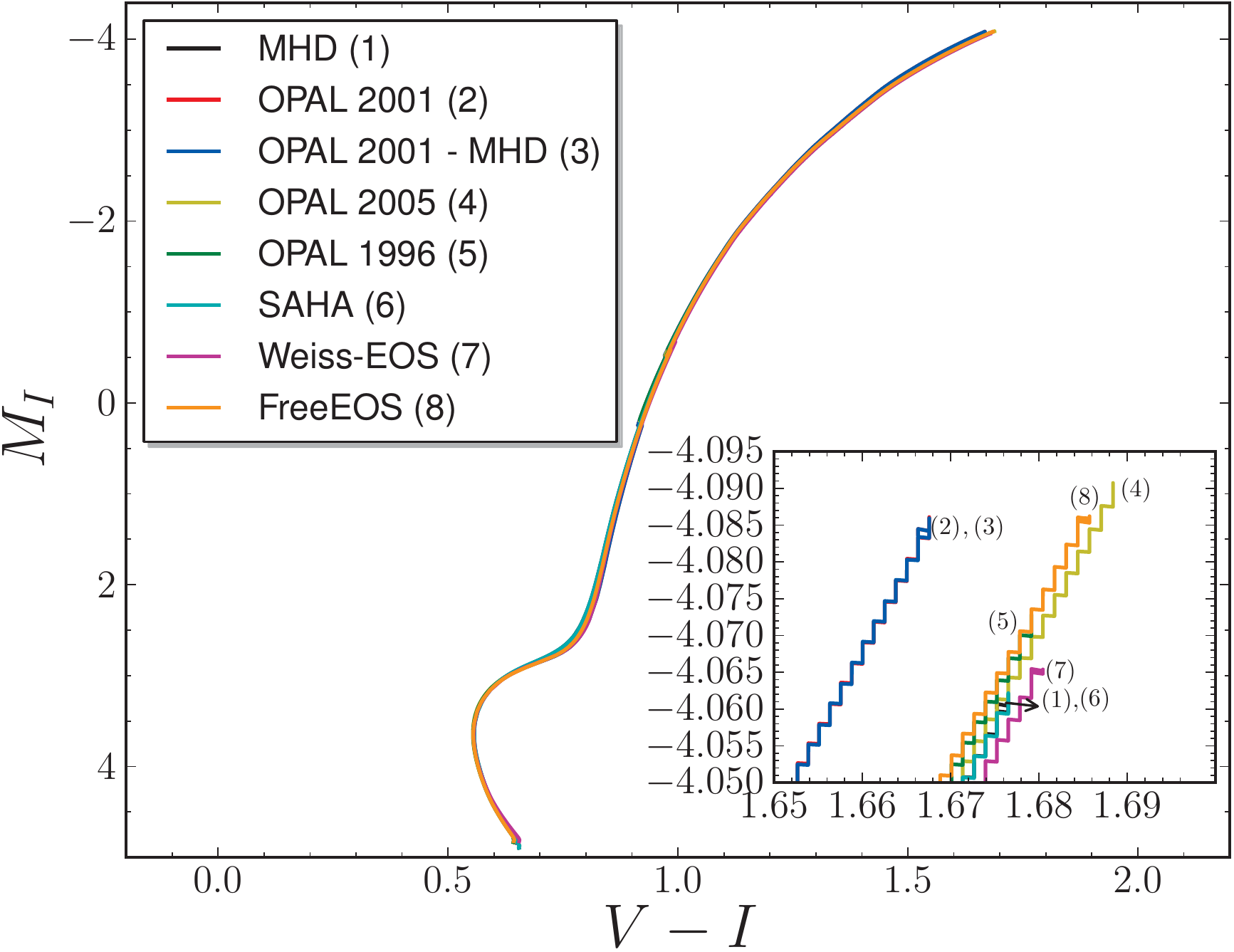}
   \caption{CMDs for different EOS based on GARSTEC.
   Inset: Numbers mark the TRGB for the
   respective EOS cases.
   The steps in the inset are due to limited number of digits in model output.}
   \label{eos}
\end{figure}

Different kinds of EOS exist in the literature that can be
implemented in stellar evolution codes. PGPUC uses FreeEOS
\citep[][see table 2 of \citeauthor{pgpuc}
\citeyear{pgpuc}]{freeEOS}, while other codes use other
prescriptions. To test impact on the TRGB we select different EOS
and compare with FreeEOS. This study was performed with the GARSTEC
stellar evolution code \citep{garstec} because in PGPUC only two EOS
have been implemented, whereas in GARSTEC we can choose between
eight different versions:
\vspace{-0.15cm}
\begin{enumerate}
\item MHD \citep{mhd}
\item OPAL 2001 \citep{opal02}
\item OPAL 2001--MHD (OPAL extended when needed with MHD)
\item OPAL 2005 (Rogers, unpublished)
\item OPAL 1996 \citep{opal96}
\item SAHA-degeneracy \citep{kip12}
\item \citet{weiss99}
\item FreeEOS
\end{enumerate}
\vspace{-0.15cm}
Figure~\ref{eos} shows the impact of these different EOS
prescriptions and the change in $M_{I,{\rm TRGB}}$. Other stellar
parameters were held fixed at our benchmark values and the same
physical input of PGPUC was used with the exception of the $T-\tau$
relation. We infer an $M_{I,{\rm TRGB}}$ uncertainty of
$-0.0045/{+}0.0242$~mag.
\subsection{Color transformations and bolometric corrections}\label{sec:bc}
\begin{figure}[!t]
  \centering
  \includegraphics[width=0.8\columnwidth]{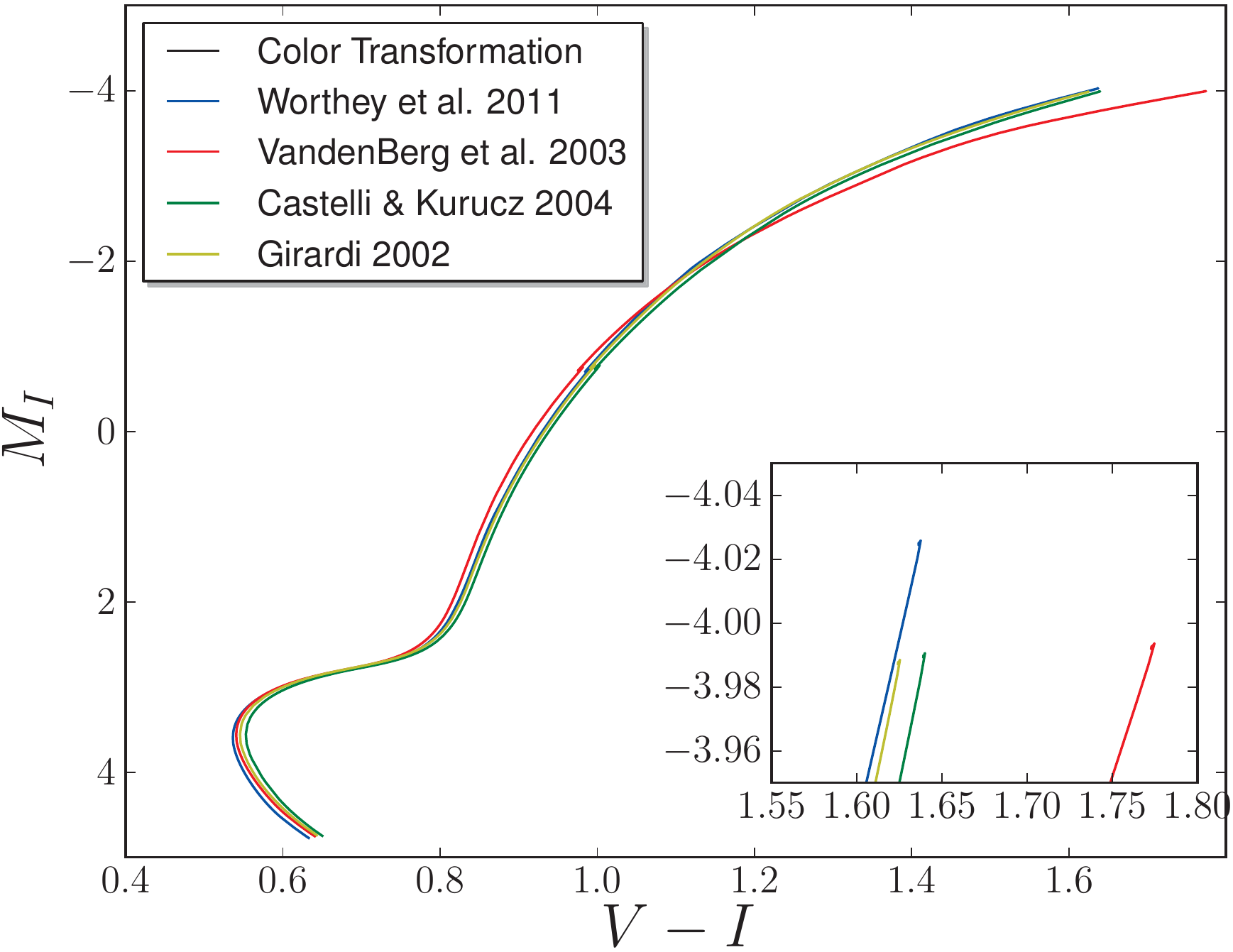}
  \caption{As in Fig.~\ref{comp1}, but for different prescriptions for the
  color transformations and bolometric corrections.}
  \label{compcol}%
\end{figure}

Different approaches can be followed to obtain the BCs and color
transformations that are needed to translate luminosities and
temperatures given by stellar evolution codes to observational
quantities. One may use purely theoretical model atmospheres,
re-calibrated model atmospheres, or empirical transformations
\citep[see, e.g,][]{S02}. Naturally, each of these routes possesses
its own uncertainties, and it is beyond the scope of this paper to
perform a critical evaluation of this complex subject. Nonetheless,
one may obtain a sense of the uncertainties involved by comparing
the prescriptions provided by different authors.
Figure~\ref{compcol} compares the predicted CMD loci, adopting the
prescriptions of \citet{castelli04}, \citet{V03},
\citet[]{girardi02}, and \citet{W11}.\footnote{Here we adopt
$M_{{\rm bol},\odot}=4.7554 \pm 0.0004$~mag, following {\tt
http://www.pas.rochester.edu/$\sim$emamajek/sun.txt}.} We see that
the differences in predicted TRGB color are larger than those in the
TRGB magnitude. As to the latter, the main discrepancies are between
the formulations of \citet{W11} and \citet[]{girardi02}.

As an error estimate caused by the BC we adopt the uncertainties
provided explicitly by \citet{W11}. For the TRGB magnitude, these
are significantly larger than the spread between the results of the
different formulations shown in Fig.~\ref{compcol}. In this sense,
using the error estimate of \citet{W11} is conservative.

Note that the error of the predicted $M_{I,{\rm TRGB}}$ value
depends strongly on $\mu_\nu$ because the effective temperature of
the TRGB decreases for higher $\mu_\nu$ values (see Fig.~\ref{hr}),
thus implying greater errors in the BC (see Table~2 of
\citeauthor{W11} \citeyear{W11}). The uncertainties extracted from
\citet{W11} for the relevant range of $\mu_\nu$ are well represented
by
\begin{equation}\label{eq:BCerror}
\sigma_{\rm BC}=(0.08+0.013\,\mu_{12})~{\rm mag}\,,
\end{equation}
where again $\mu_{12}=\mu_\nu/10^{-12}\mu_{\rm B}$. Note that this
adopted error is more likely to be seen as an estimated maximum
error and not a 1-$\sigma$ uncertainty.

\subsection{Comparison with other stellar evolution codes}

In Fig.~3 of \citet{pgpuc}, a comparison between PGPUC and other
stellar evolution codes is made, as far as predicted CMD positions
are concerned. This figure implies that PGPUC reaches the TRGB with
a luminosity that is lower than predicted by other codes. Our
discussion of different EOS cases in Sect.~\ref{sec:eos}, calculated
with GARSTEC, reveals that for identical input parameters and the
same EOS, PGPUC finds $M_{I,{\rm TRGB}}$ about 0.05~mag dimmer than
GARSTEC. It would be too difficult to isolate the impact of the
numerical algorithms and different input physics adopted by the
different authors upon this result, but the reliability of the
numerical approach adopted in the PGPUC code was discussed at some
length by \citet{as94}. Rather than associating a formal ``error
bar'' stemming from this comparison, we simply emphasize that the
$\mu_\nu$ constraints obtained using PGPUC are less stringent (more
conservative) than those that would have been derived using most
other evolutionary codes.

\subsection{Summary}\label{sec:theory-sum}

The sources of theoretical error are summarized in
Table~\ref{tab:error-theory}, except for the BC, given in
Eq.~(\ref{eq:BCerror}). All of these uncertainties are systematic
(not statistical) and constitute our best estimates of the maximum
error. The associated probability distributions are in most cases
completely unknown, so we make the simplest possible choice and use
top-hat, flat probability distributions in the ranges shown in
Table~\ref{tab:error-theory}. The convolution of all these
distributions is well represented by a Gaussian with mean and
standard deviation
\begin{eqnarray}
\label{eq:shift}
\langle \delta M^{\rm the}_I\rangle &=& 0.039 \, \rm mag \, , \\
\sigma_{\rm the} &=& \sqrt{0.039^2+(0.046+0.0075\mu_{12})^2} \, \rm mag \ .
\end{eqnarray}
The shift of the mean is due to the fact that the error ranges are
not symmetric (and we have chosen flat probabilities).
Otherwise our procedure corresponds to adding systematic
errors in quadrature as recommended, for example, by
\citet{rbea02}.

One important contribution to the shift comes from mass-loss. Our
theoretical prediction $M_I^{\rm the}$ of Eq.~(\ref{eq:DeltaMI}) was
computed setting zero mass-loss,  but we have seen that mass-loss
falls in the range 0.12--$0.28\,M_\odot$. Since $M_I^{\rm the}$
increases with the mass loss (the TRGB becomes dimmer) our best
estimate has to be shifted accordingly (to $0.2$ $M_\odot$, which
corresponds to 0.03~mag for $M_{I,{\rm TRGB}}$).  For a single
top-hat distribution, the standard deviation is $R/\sqrt{3}$ where
$R$ is the half-width. The standard deviation of the convolution,
$\sigma_{\rm the}$, is well represented by $\sqrt{\sum_{i=1}
R^2_i/3}$ as it corresponds to the convolution of a sum of
uncorrelated errors. This completes our estimate of theoretical
uncertainties.

\begin{table}[h]
\centering
\caption{Error budget in theoretically predicted $M^{\rm the}_{I,{\rm
TRGB}}$}
\begin{tabular}{lll}
\hline\hline
\label{tab:error-theory}
\textbf{Input quantity}&\textbf{Adopted Range}&\textbf{\boldmath{$\Delta M_{I,{\rm TRGB}}~[\textbf{0.01~mag}]$}}\\
\hline
Mass ($M_\odot$)          & $0.820\pm0.025$       & $\pm 0.2$\\
$Y$                       & $0.245\pm0.015$       & $\pm 1.0$\\
$Z$                       & $0.00136\pm0.00035$   & $+0.7$/$-0$\\
${\rm [\alpha/{\rm Fe}]}$ & $0.3\pm0.1$           & $\mp 0.4$\\
$\alpha_{\rm MLT}$ & $\alpha_{\rm MLT}^{\rm calibrated}\pm0.2$ & $\pm 5.6$\\
Atomic diffusion          & See text              &  $+0$/$-0.6$\\
Boundary conditions       & $(1\pm0.05)\,T(\tau)$ & $\mp 0.7$\\
$\kappa_{\rm rad}$        & $\pm10\%$             & $\mp 0.02$\\
$\kappa_{\rm c}$          & $\pm10\%$             & $\pm 1.6$\\
Nuclear Rates                  & See Table~\ref{table:1}& $\pm 1.9$\\
Nuclear Screening      & $\pm20\%$             & $\pm 1.1$\\
Neutrino emission         & $\pm5\%$              & $\mp 1.3$\\
EOS                       & 8 cases               & $+2.4$/$-0.5$\\
Mass loss ($M_\odot$)     & 0.12--0.28            & $+2.2/+3.5$\\
\hline
\hline
\end{tabular}
\end{table}

\section{Comparing observational and theoretical results}\label{sec:comp}

After all ingredients are complete, we can now compare the predicted
and observed $I$-band absolute brightness of the TRGB in M5. For the
empirical TRGB brightness we have found the value of
Eq.~(\ref{eq:photoerror}). This result is to be compared with the
predicted value, Eq.~(\ref{eq:DeltaMI}), corrected for mass-loss and
other non-symmetric systematics according to Eq.~(\ref{eq:shift}),
i.e.,
\begin{eqnarray}
\nonumber
M^{\rm the}_{I,\rm TRGB}
&=& -3.99-0.23\,\left(\sqrt{\mu_{12}^2+0.64}-0.80-0.18\,\mu_{12}^{1.5}\right) \\
&& \pm \sqrt{0.039^2+(0.046+0.0075\mu_{12})^2} .
\end{eqnarray}
In the absence of neutrino dipole moment ($\mu_{12}=0$), this corresponds
to $M^{\rm the}_{I,\rm TRGB}=-3.99\pm 0.07$. 

\begin{figure}[t]
  \centering
  \includegraphics[width=1.0\columnwidth]{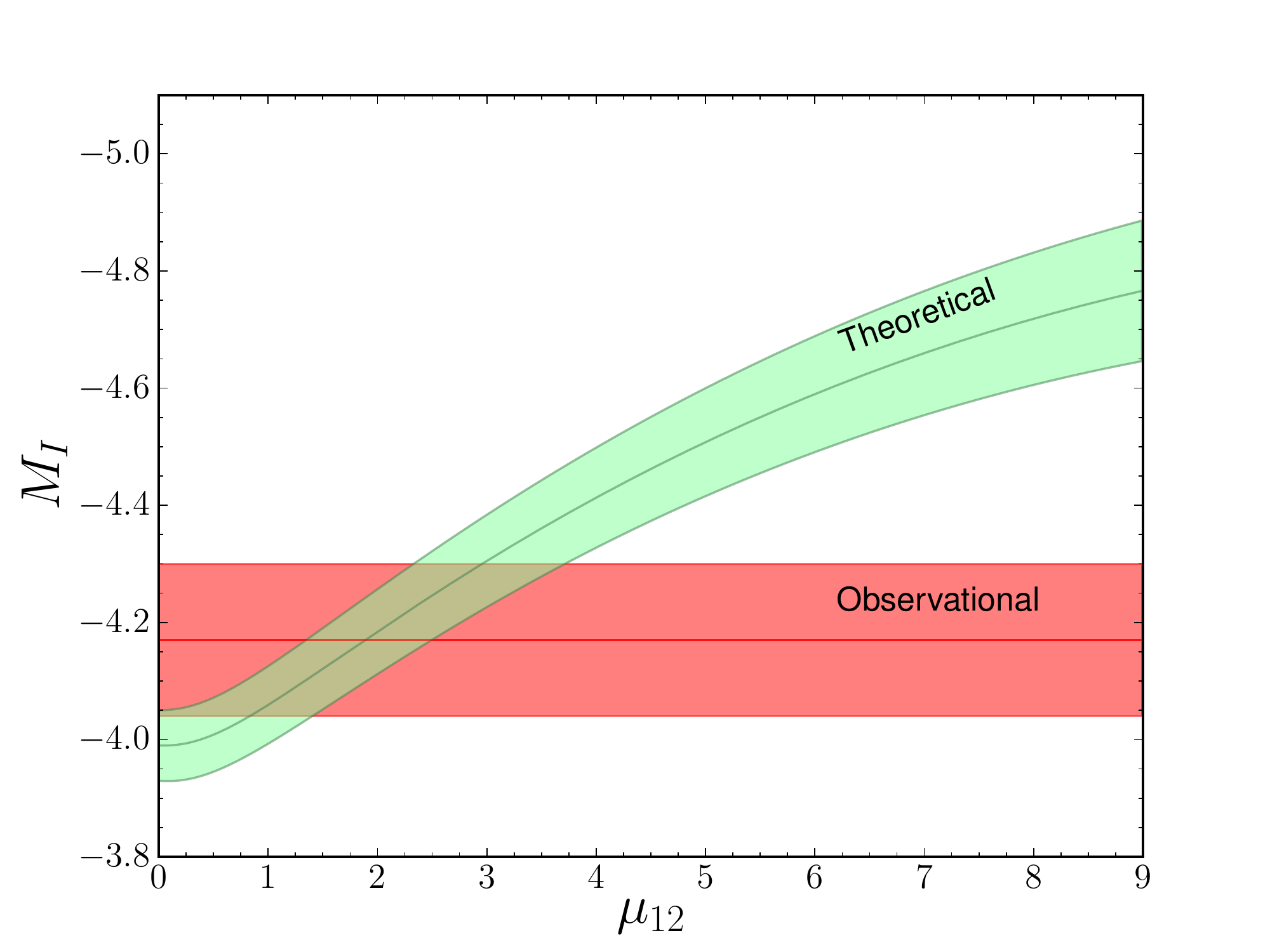}
  \caption{Absolute TRGB $I$-band magnitude inferred from
  observations and predicted by our calculations, both with $1\,\sigma$ error bands.}
  \label{fig:constraint}
\end{figure}

Let us now discuss the implications of a non-zero neutrino dipole moment. 
Figure~\ref{fig:constraint} shows the predicted
and observed values for $M_{I,{\rm TRGB}}$ (with $1\, \sigma$ errors)
as a function of $\mu_{12}$. 
In order to compare them, we combine the observational and theoretical 
errors in quadrature. 
The most robust information we can extract from our study are upper limits on $\mu_{12}$. 
Integrating the combined probability distribution from 
$\mu_{12}=0$ to the limiting value, we obtain the following upper limits, 
\begin{eqnarray}\label{eq:constraints}
\mu_{\nu} &<&2.6 \times 10^{-12}\mu_{\rm B}\quad \hbox{at 68\% CL,}
\nonumber\\
\mu_{\nu} &<&4.5\times 10^{-12}\mu_{\rm B}\quad \hbox{at 95\% CL.}
\end{eqnarray}
In Fig.~\ref{fig:constraint} we see a small disagreement at 
$\mu_\nu=0$, which naively suggests a best fit value of $\mu_{12}\sim 2$.
This small discrepancy can be blamed entirely on the distance uncertainty, 
and implies that we obtain a relatively poor limit. 
A more precise future distance determination will lead either to a much 
improved limit or reveal significant deviations from standard stellar 
evolution theory. 

\section{Conclusions}\label{sec:conclusions}
We have derived the new astrophysical constraints on neutrino dipole
moments $\mu_\nu$ shown in Eq.~(\ref{eq:constraints}). We have used
precise observations of the GC M5, carefully determined its TRGB
position, compared against theoretical predictions for different
$\mu_\nu$ values, and most importantly have taken into detailed
account the different sources of error affecting both the empirical
and theoretical TRGB position.

Our 68\%~CL limit is almost identical with the constraints found in
the earlier literature by similar methods (\citeauthor{R90}
\citeyear{R90}, \citeauthor{RW92} \citeyear{RW92}, \citeauthor{Ca96}
\citeyear{Ca96}), although it is not possible to associate a formal
confidence limit to the older results. In this sense, our new result
is more robust because it is based on state-of-the-art astronomical
data, evolutionary calculations using up-to-date input physics, and
a careful analysis of the several different error sources affecting
the TRGB position, both observational and theoretical.

The paucity of stars on the upper RGB is not our main source of
error, yet repeating our analysis for additional GCs should allow us
to improve our limits and check for overall consistency in other
cases. The systematic errors associated with the theoretical
prediction is dominated by the uncertainty of the BC, followed by
the mixing-length theory and uncertain mass loss. A better
understanding of the BC uncertainty, in particular, would strongly
help to improve our results. The main observational error, as well
as the largest single source of uncertainty, derives from the
distance modulus which should be improved in future by the results
of the GAIA mission. Depending on the distance found and the
remaining errors, a significant improvement of our results could be
obtained.

The stellar energy-loss limit remains the most restrictive
constraint on $\mu_\nu$ (valid for only neutrinos with masses below the keV range). The most restrictive laboratory limit uses
the $\bar\nu_e$ flux from reactors and studies the electron recoil
spectrum upon $\bar\nu_e$ scattering, leading to the constraint
$\mu_{\bar\nu_e}<32\times10^{-12}\,\mu_{\rm B}$ (90\%~CL) on
neutrino magnetic or transition moments that are connected to
$\bar\nu_e$ \citep{beda2010}. This quantity is different from our
$\mu_\nu$ which effectively sums over all direct and transition
moments between all flavors and therefore is more general. It also
applies to transition moments between ordinary active and putative
sterile neutrinos, provided the latter are light enough to be
emitted from the degenerate helium core near the TRGB, i.e., the
mass is safely below the relevant plasma frequency of about
10--20~keV.

Neutrino dipole moments also lead to radiative neutrino decays of the
form $\nu_2\to\nu_1+\gamma$, allowing one to derive constraints on
$\mu_\nu$ from the absence of anomalous photon fluxes in connection
with astrophysical neutrino fluxes. However, neutrino oscillation
experiments now tell us that the mass differences between ordinary
active neutrinos are very small, suppressing radiative decays by
phase-space effects, so that the stellar energy loss argument is more
constraining \citep{Raffelt:1998xu}.

Since our $1\sigma$ limit is similar to previously stated GC bounds
on $\mu_\nu$, we expect that other previous constraints, for example
on the axion-electron interaction, remain valid if interpreted as
$1\sigma$ limits. We will study other cases more explicitly in a
forthcoming paper.

GCs remain powerful particle-physics laboratories. It is timely to
put such results on a firmer observational basis and understand more
deeply the associated uncertainties and errors.

\begin{acknowledgements}
The authors are grateful to S.~Cassisi for useful discussions and
for kindly providing unpublished evolutionary tracks computed using
BaSTI. Support for N.V.\ and M.C.\ is provided by the Chilean
Ministry for the Economy, Development, and Tourism's Programa
Iniciativa Cient\'{i}fica Milenio through grant P07-021-F, awarded
to The Milky Way Millennium Nucleus; by Proyecto Fondecyt Regular
\#1110326; by the BASAL Center for Astrophysics and Associated
Technologies (PFB-06); and by Proyecto Anillo ACT-86. Support for
N.V.\ is also provided by MECESUP Project No. PUC0609 (Chile).
G.R.\ acknowledges partial support by the Deutsche
Forschungsgemeinschaft through grant No. EXC 153 and by the European
Union through the Initial Training Network ``Invisibles,'' grant
No.\ PITN-GA-2011-28944. J.R.\ acknowledges support by the Alexander
von Humboldt Foundation.
\end{acknowledgements}

\bibliographystyle{aa}
\bibliography{nviauxM5}

\end{document}